\RequirePackage{fix-cm}
\documentclass[smallextended]{svjour3}       
\smartqed  
\usepackage{graphicx}
\usepackage{natbib}
\usepackage{txfonts}
\begin{document}

\title{Against all odds? Forming the planet of the HD196885 binary
}

\titlerunning{Planet formation in HD196885}        

\author{P. Thebault
}


\institute{P. Thebault \at
LESIA, Observatoire de Paris,
F-92195 Meudon Principal Cedex, France \\
              \email{philippe.thebault@obspm.fr}           
}

\date{Received: date / Accepted: date}

\maketitle

\begin{abstract}

HD196885Ab is the most "extreme" planet-in-a-binary discovered to date, whose orbit places it at the limit for orbital stability. The presence of a planet in such a highly perturbed region poses a clear challenge to planet-formation scenarios. We investigate this issue by focusing on the planet-formation stage that is arguably the most sensitive to binary perturbations: the mutual accretion of kilometre-sized planetesimals. To this effect we numerically estimate the impact velocities $dv$ amongst a population of circumprimary planetesimals. We find that most of the circumprimary disc is strongly hostile to planetesimal accretion, especially the region around 2.6AU (the planet's location) where binary perturbations induce planetesimal-shattering $dv$ of more than 1km.s$^{-1}$. 
Possible solutions to the paradox of having a planet in such accretion-hostile regions are 1) that initial planetesimals were very big, at least 250km,  2) that the binary had an initial orbit at least twice the present one, and was later compacted due to early stellar encounters, 3) that planetesimals did not grow by mutual impacts but by sweeping of dust (the "snowball" growth mode identified by Xie et al., 2010b), or 4) that HD196885Ab was formed not by core-accretion but by the concurent disc instability mechanism. All of these 4 scenarios remain however highly conjectural.

\end{abstract}

\section{Introduction} \label{intro}

\subsection{Planets in binaries}

Studying planet formation in binaries is of fundamental importance, as a majority of main sequence stars are members of multiple systems \citep{duq91}. Moreover, planets in binaries are no longer theoretical concepts, as about 20\% of all known exoplanets have been found to inhabit multiple stellar systems \citep{desi07,mug09}. Most of the planet-bearing binaries have large separations, often in excess of 1000\,AU, for which the influence of the companion star on the planet region, and the planet formation process, is probably limited. However, a handful of planets inhabit much tighter binaries, which separations as small as $\sim\,20\,$AU: Gl86, with a planet at 0.11AU and a companion at 18.4\,AU \citep{quel00,lag06}, HD 41004 with a planet at 1.64\,AU and companion at 23\,AU \citep{zuck04} and $\gamma$ Cephei, for which the planet lies at 2.04\,AU and the companion at 20.18\,AU \citep{hatz03,neuh07}.
For these systems (especially 41004 and $\gamma$ Cephei), the closeness of the companion must have had an influence on the way the planet formed and dynamically evolved. 
Studying how such planets came about is of great interest, all the more because close-binaries can be used as a test bench for planet formation models, by confronting them to an unsual environment where some crucial parameters might be pushed to extreme values.

Historically, the first issue that has been investigated is that of the long term stability of planetary orbits in binaries. The reference work on this issue remains probably that of \citet{holw99}, who derived empirical expressions for orbital stability as a function of binary semi-major axis $a_B$, eccentricity $e_B$ and mass ratio $\mu$. Later studies have shown that, reassuringly, all known exoplanets in multiple systems are on stable orbits \citep[e.g.][]{dvo03,hagh10}, although the case for HD41004 is not fully settled yet, as it depends on the yet unconstrained eccentricity of the binary orbit \citep{hagh10}.

\subsection{Planet formation in binaries}

The next step is to study under which condition such planets can $form$, as the constraints for planet formation might be very different from those for orbital stability. This issue is a much more difficult one. Planet formation is indeed a very complex process, believed to be the succession of several stages \citep[e.g.][]{liss93}, each of which could be affected in very different ways by the perturbations of a secondary star.
Not surprisingly, the effect of binarity on each of these different stages is usually investigated in separate studies. 

The initial phase of planet formation, i.e., the formation and evolution of a gaseous protoplanetary disc, has been investigated early on by \citet{arty94} and \citet{savo94}, who have shown that the circumprimary disc is tidally truncated by the companion. This truncation occurs at a location comparable to the outer limit for dynamical stability and thus safely lies beyond the position of all detected exoplanets. It could nevertheless pose a problem, especially for giant planet formation, as it deprives the disc from a large fraction of its mass. For the specific case of $\gamma$ Cephei, however, \citet{jang08} have found that there is probably enough mass left in the truncated disc to form the observed giant planet. But another, potentially more troublesome consequence of disc truncation is that it shortens the viscous lifetime of the disc, and thus the timespan for gaseous planet formation. This effect seems to have been observationally confirmed by \citet{ciez09}, who found that young binaries with separation $\leq 100\,$AU have a lower probably of hosting circumstellar dust in the innermost few AU around each star \citep{duch10}, even if some close binaries do show signs of a hot circumprimary disc.

The next stage of planet formation, the condensation of small grains and their growth into larger pebbles and eventually kilometre-sized planetesimals, has not been extensively studied in the context of binary systems. One main reason is probably that this stage is the one that is currently the least understood even in the "normal" context of single stars \citep[e.g.][]{blum08}, so that extrapolating it to perturbed binaries might seem premature. A noteworthy exception is the study by \citet{nels00} showing that for an equal-mass binary of separation 50\,AU, temperatures in the disc might stay too high to allow grains to cendense. But, as ackowledged by the author himself, these results are still preliminary and a full study of this issue has yet to be undergone. Very recently, \citet{zsom10} showed that, even if grains can condense, binary perturbations might impend their growth by mutual sticking because of too high impact velocities. 

The stage for which the influence of a companion is probably best understood is the final step of planetary accretion, leading from Lunar-sized embryos to fully formed planets. Several studies have shown that the regions where embryo accretion can proceed roughly correspond to those for orbital stability \citep{barb02,quin07,gued08,hagh10}. This is a further reassuring result for all known exoplanets-in-binaries, in particular the archetypal case that is $\gamma$ Cephei, for which the extensive study of \citet{kley08} showed that, $if$ embryos can form around 2\,AU from the primary, then they can evolve to form planetary cores at the present planet location.

\subsection{Planetesimal accretion in binaries: latest results} \label{planaccr}

The stage that has been the most extensively studied in recent years is the one just before the final embryos-to-planets phase, i.e., the one leading, through mutual accretion of kilometer-sized planetesimals, to the embryo themselves. The reason for this intense research activity is that this stage is potentially the one that is most affected by binary perturbations. Indeed, in the standard planet-formation version, this stage proceeds through fast runaway and oligarchic growth that require very low impact velocities between colliding bodies, typically smaller than their escape velocity, i.e., just a few m.s$^{-1}$ for kilometer-sized objects \citep[e.g.,][]{liss93}. It is thus very sensitive to dynamical perturbations, the crucial parameter sealing the fate of the planetesimal population being the distribution of {\it encounter velocities} $dv$ among them. As this field of research is a (very) fast evolving one, let us here briefly summarize and synthesize the main results obtained so far, especially in the past couple of years.

Early works \citep[e.g.][]{hepp78} revealed that estimating the $dv$ distribution is not straightforward. In the complex dynamical environment of a binary, $dv$ is indeed no longer directly proportional to the bodies eccentricity $e$ and inclination $i$, because planetesimal orbits are strongly phased and not randomly distributed in periastron and ascending nodes. 

The pioneering study of \citet{marz00} later showed that a fundamental mechanism controlling planetesimal dynamical evolution is the coupling between secular perturbations forced by the companion and friction with the primordial gas that is left in the protoplanetary disc at this stage. This coupling results in a strong phasing of planetesimal orbits. \citet{theb04,theb06} have shown that this phasing is size-dependent, so that $dv$ are small between equal-sized objects but can reach very high values for bodies of different sizes. For most "reasonable" size distributions within the planetesimal population, the differential phasing effect is the dominant one \citep{theb06}.
This can lead to an accretion-hostile environment in vast regions of the circumprimary disc, which can stretch much closer to the primary than the radial limit for orbital stability or embryo accretion. Furthermore, even in most of the regions where planetesimal accretion $is$ possible, it cannot proceed in the same way as around a single star, because the $dv$ increase is still enough to strongly slow down and impede the runaway growth mode. A worrying result was that, for the emblematic $\gamma$ Cephei case, the location at which the planet is observed is probably too perturbed to allow for the planetesimal accretion stage to proceed \citep{paard08}. Similarly troublesome results were later obtained for potential planets in the habitable zone (HZ) of both stars of arguably the most famous binary system: $\alpha$ Centauri \citep{theb08,theb09}.

These pessimistic studies were all considering a simplified static and axisymetric gas disc. However, the simulations of \citet{paard08} showed that the situation gets even worse when considering a dynamically evolving gas disc: gas streamlines and planetesimals follow very different orbits, which in most cases increases gas friction and thus the accretion-hostile effect of differential phasing of planetesimals. 
On a more positive note, \citet{xie09} and \citet{xie10} showed that a small inclination of a few degrees between the circumprimary gas disc and the binary orbital plane could in fact help accretion. This is because planetesimal orbital inclinations are segregated by size, thus favouring low-$dv$ impacts between equal-sized bodies over high-$dv$ impacts between differently-sized objects. As a result, planetesimal accretion could become possible, albeit in a very slowed down form, in the HZ of $\alpha$ Cen. 
However, a very recent study by \citet{frag11} seems to indicate that taking into account the effect of the gas disc's gravity could offset this positive effect of orbital inclination, leading to high $dv$ dynamical environments.

This planetesimal accretion issue is thus far from having been solved, and is in any case very dependent of the set up: binary orbit, relative masses of the star, location in the circumprimary disc, etc. \citep{theb06}.

\subsection{HD196885}

Given the huge parameter space to explore, most studies of planet formation in binaries consider only one specific and illustrative example. It is usually either $\alpha$ Centauri, because of its obvious interest as our closest neighbour and potential for planet detection \citep{gued08}, or $\gamma$ Cephei, because it was until recently the most "extreme" exoplanet in a binary, in the sense that it is the one, given its large distance to the primary and proximity to the secondary, for which companion perturbations are expected to be the strongest.

This "privileged" position of $\gamma$ Ceph has however been challenged by a recent study by \citet{chau10} on HD196885. This system had already been identified as a planet-hosting binary \citep{chau07}, but only the companion's projected distance was known, not the binary's orbit. With new observations, \citet{chau10} were able to constrain this orbit to $a_B =21.0 \pm 0.86\,$AU and $e_B=0.42 \pm 0.03$, and refined the exoplanet orbit to $a_P=2.6 \pm 0.1$ and $e_P= 0.48 \pm 0.02$ \citep[a value close to the earlier esimate by][]{corr08}. This makes HD196885Ab a more extreme binary exoplanet than $\gamma$ Cephei Ab. One way to quantify it is by looking at the \citet{holw99} criteria for orbital stability, which gives a critical radial distance (to the primary) of $a_{crit}=3.71\,$AU for $\gamma$ Ceph and $3.82\,$AU for HD196885, thus placing the planet at $\sim 1.7\,$AU from the orbital stability limit in $\gamma$ Ceph but at only $\sim 1.2\,$AU for HD196885.
In fact, this planet's eccentricity, $e_P=0.48$, makes it reach $\sim 3.85\,$AU, i.e. slighly $beyond$ $a_{crit}$, although this does not necessarily mean its orbit is unstable, since $a_{crit}$ is not a razor sharp boundary (the formula of \citet{holw99} comes with error bars delimiting a "gray" area around $a_{crit}$).
However, a preliminary numerical analysis by \citet{chau10} indicates that this planet might possibly be on an unstable orbit, unless there is a high (unconstrained) inclination between the binary plane and the planet orbit. These stability issues need to be further investigated, but it is clear that this system is by far the most "extreme" planet-in-a-binary so far.

\subsection{present work}

We shall not reinvestigate the issue of the planet's long-term stability, nor shall we investigate the possible cause for its high eccentricity of 0.48. We shall here focus on the key issue of the $formation$ of a planet in such a perturbed environment. HD196885Ab is indeed clearly the planet that poses the strongest challenge to any planet-formation model, especially regarding the planetesimal-accretion stage.
We numerically investigate under which conditions this stage might, or might not, proceed in HD196885A's circumprimary disc. As with most previous similar studies, we follow the distribution of one crucial parameter: impact velocities amongst planetesimals. Our aim is to identify the regions where the dynamical environement is too perturbed to allow classical core-accretion of a planet. 
To obtain conservative results, we shall consider the most favourable (i.e., accretion friendly) assumptions for our simulation: axisymetric static gas disc, no self gravity. We present our numerical model in Sec.\ref{model}.  Our main results are presented in Sec.\ref{results}. In Sec.\ref{discu}, we discuss the robustness of our main conclusions, i.e., that most of the circumprimary disc is hostile to accretion, and consider several possible solutions to the apparent paradox of having a planet in a accretion-hostile region. Conclusions are presented in Sec.\ref{conclu}.

\section{Model}\label{model}

We follow a population of $N=2\times 10^{4}$ planetesimals, sampling a much larger population of "real" physical planetesimals, orbiting the primary and dynamically perturbed by the companion. The forces acting on the particles are both stars' gravity and friction with the primordial gas disc. The code has a built-in collision search algorithm, tracking, at each time step, all mutual encounters between all bodies. In order to yield a statistically significant number of encounters despite the limited number of particles as compared to a real planetesimal population, we resort to the usual method of assigning an inflated radius to each particle \citep[e.g.][]{theb98,marz00,char01,lith07,xie08}. For a more detailed description of our algorithm, see for example \citet{theb98} and \citet{theb06}.

Gas drag is computed following \citet{wd85}:
\begin{equation}
        \vec{F}  =  - K v_{p-g} \vec{v_{p-g}}  , 
\end{equation}
where $\vec{F}$ is the force per unit mass, $\vec{v_{p-g}}$ the velocity of the planetesimal with respect to the gas, $v_{p-g}$ the velocity modulus, and $K$ is the drag parameter given by:
\begin{equation}  
        K  =  \frac{3 \rho_{\rm g} C_d} {8 \rho_{{\rm pl}} s_{}}  ,
\end{equation}  
where $\rho_{\rm g}$ is the gas density, and $\rho_{{\rm pl}}$ and $s$ are the planetesimal density and radius, respectively. The coefficient $C_d$ is a dimensionless quantity related to the shape and size of the body ($\simeq 0.4$ for spherical bodies).  
The gas disc is assumed to be static and axisymetric. For the gas density and gas velocity, we follow \citet{take02} and assume
\begin{equation}
\rho_g(r,z)=\rho_{g0}\left(\frac{r}{AU}\right)^p {\rm exp}\left(-\frac{z^2}{2h_g^2}\right),
\end{equation}
and
\begin{equation}
v_g(r, z)=v_{k, {\rm mid}}\left[1+\frac{1}{2}\left(\frac{h_g}{r}\right)^2 \left(p+q+\frac{q}{2}\frac{z^2}{h_g^2}\right)\right]
\end{equation}
where $h_g(r) = h_0( r/ {\rm AU} )^{(q+3)/2}$ is the scale height of the gas disc and $v_{k, {\rm mid}}$ is the Keplerian velocity in the midplane. We consider the Minimum Mass of Solar Nebula \citep[MMSN, see][]{haya81} as a reference disc, where $p = -2.75$, $q = -0.5$, $\rho_{g0} =1.4 \times 10^{-9}{\rm g. cm^{-3}}$, and $h_0 = 4.7 \times 10^{-2}$\,AU.

The static and axisymetric assumption are taken for computing time reasons but is of course a crude simplification of the real behaviour of a circumprimary gas disc, which should react to the companion's perturbations and display pronounced eccentric shapes and azimutal anisotropies \citep[e.g.][]{arty94}. However, preliminary studies with evolving gas discs \citep{paard08} have shown that gaseous friction, and in particular the differential phasing effect according to planetesimal size, is higher than for a static gas disc.
This is because the gas disc gets eccentric \citep{good06}, with an eccentricity $e_g$ and a precession rate that in most cases strongly departs from that of the planetesimals \citep[see for instance a clear illustration in Figs.9 and 10 of][]{paard08}. This increases the relative velocities between planetesimals and gas streamlines, and thus the friction of the latter on the former.
This makes the systems globally more accretion hostile than in the axisymetric case (see Sec.\ref{planaccr}). As a consequence, the axisymetric assumption should be regarded as a limiting best-case scenario for planetesimal accretion.

\subsection{setup} \label{setup}

\begin{table}
\caption[]{Setup for the nominal run (see main text for parameter definition)}
\label{setupnom}
\begin{tabular}{ll}
\hline
Binary mass ratio & $\mu = 0.35$\\
\qquad \quad semi-major axis & $a_b = 21.0$\\
\qquad \quad eccentricity & $e_b = 0.42$\\
Number of test particles& $2\times 10^{4}$\\
Inflated radius (collision search routine)& $5\times 10^{-5}$\,AU\\
Physical radius & 1\,km$\leq s \leq$10\,km\\
Initial semi-major axis & 0.9\,AU$\leq a \leq$3.1\,AU\\
Initial eccentricity & $0\leq e \leq 5\times 10^{-5}$\\
Initial inclination & $0\leq i \leq 2.5\times 10^{-5}$\\
Gas Disc density at 1AU & $\rho_{0}=1.4\times 10^{-9}$g.cm$^{-3}$\\
\qquad \qquad radial profile & $\rho_g(r) \propto r^{-2.75}$\\
\qquad \qquad scale heigh & $h_g = 4.7 \times 10^{-2} (r/$1AU$)^{1.25}$\,AU\\
\qquad \qquad vertical profile & $\rho_g(z) \propto \,$exp$(-z^{2}/2h_g)$\\
\hline
\end{tabular}
\end{table}

We consider a disc of planetesimals with initial semi-major axis $0.9 \leq a \leq 3.1\,$AU, having randomly distributed orbits (longitude of periastron and of ascending node) and initial eccentricity $0 \leq e \leq 5 \times 10^{-5}$ and inclination $0 \leq i \leq 2.5 \times 10^{-5}$. This ensures that initial encounter velocities are such as $dv_{init} \sim 1-2$m.s.$^{-1}$, approximately the escape velocity of a 1km body. This is the velocity distribution expected in an unperturbed population of km-sized planetesimals, in which runaway growth can proceed \citep{liss93}.
The planetesimals physical sizes are randomly distributed between 1 and 10\,km. The inflated radius for the collisional search routine is $5\times 10^{-5}$\,AU. This size is large enough to yield a statistically significant number of impacts, but small enough not to introduce any bias in estimating $dv$.
The set-up for our nominal run is summarized in Tab.\ref{setupnom}.

\subsection{Accretion and Fragmentation Prescription} \label{acpresc}

Our simulations provide us with the distribution of encounter velocities $dv_{(s1,s2)}$ for all impacting planetesimal pairs of sizes $s_1$ and $s_2$. A key issue is then to interpret these velocities and see if they are low enough to allow accretion or are on the contrary too high and lead to mass loss or fragmentation of the impactors. Three different regimes are possible, defined by two critical velocities $v_{esc(s1,s2)}$ and $v_{ero(s1,s2)}$:
\begin{itemize}
\item $dv_{(s1,s2)} \leq v_{esc(s1,s2)}$: "Unperturbed" case. The impact velocity is only marginally increased with respect to its initial $dv_{init} \sim 1-2$m.s$^{-1}$ value and stays below the escape velocity $v_{esc(s1,s2)}$ for the impacting pair. In this case, "normal", single-star like runaway accretion is possible.
\item $v_{esc(s1,s2)} \leq dv_{(s1,s2)} \leq dv_{ero(s1,s2)}$: "Perturbed accretion" case. The impact velocity is increased beyond the escape velocity. This switches off the runaway growth mode. However, $dv$ stays at a value small enough to allow some accretion between the impacting bodies.
\item $v_{ero(s1,s2)} \leq dv_{(s1,s2)}$: "Erosion". In this case velocities are too high to allow accretion. Each impacts results in mass loss, i.e., erosion or fragmentation of one or both impactors.
\end{itemize}

While the value of $v_{esc(s1,s2)}$ is easy to derive, the erosion threshold velocity $v_{ero(s1,s2)}$ is much more difficult to estimate, as it depends on many physical parameters (particle sizes, compositon, impact velocity and angle, etc...) and because, even  for the same set of parameters, there exists many diverging estimates of the accretion/erosion limit \citep[see the discussion in][]{theb07}. 
In previous papers \citep{theb06,theb08,theb09} we considered a complex, and rather cumbersome prescription for $v_{ero(s1,s2)}$, trying to connect several possible impact regimes (cratering, shattering) and to synthesize several available estimates from the litterature. In the meantime, a remarkable study by \citet{stew09} has been published, presenting an innovative and simplified criteria for the disruption of planetesimals. Even if this study's title claims that it is only valid for the "catastrophic" disruption of planetesimals, it is in fact applicable to a larger domain, including impacts usually described as non-catastrophic (where the biggest remaining fragment is more than half the mass of the impactor). Although the results of \citet{stew09} have their limitations, we follow \citet{frag11} and adopt their prescription here because of its simplicity and its self-consistency. Its main parameter is the velocity-dependent "reduced" catastrophic disruption specific energy
\begin{equation}
Q_{RD}^{*} = q_S s_{c}^{9\alpha/(3-2\Phi)}dv^{2-3\alpha}\,\,+\,\,q_g s_{c}^{3\alpha}dv^{2-3\alpha}
\label{QRD}
\end{equation}
where $s_c=(s_{1}^{3}+s_{2}^{3})^{1/3}$ is the reduced radius of the combined projectile and target mass, and $\alpha$ and $\Phi$ are material properties. If $Q_R=0.5m_1 m_2dv/(m_1+m_2)^{2}$ is the reduced kinetic energy, then the mass $m_{lr}$ of the largest remaining fragment is given by
\begin{equation}
\frac{m_{lr}}{(m_1 + m_2)} = 1 - 0.5 \frac{Q_R}{Q_{RD}^{*}}
\label{Mlr}
\end{equation}

With the convention that $m_1 \geq m_2$, the criteria for accretion is then $m_{lr} \geq m_1 $, which translates into

\begin{equation}
dv \leq  v_{ero(s1,s2)}=\left[ 4 \left( 1+\frac{m_2}{m_1}\right)Q_{RD}^{*}\right]^{0.5}
\end{equation}

We follow \citet{stew09} and consider two limiting cases: $v_{ero1}$ for weak rocky aggregates ($\alpha = 0.4$, $\Phi = 7$, $q_S=500$, $q_g=10^{-4}$, in cgs units) and $v_{ero2}$ for strong compact rocks ($\alpha = 0.5$, $\Phi = 8$, $q_S=7\times 10^{-4}$, $q_g=10^{-4}$).

\section{Results} \label{results}

\subsection{Gas Free Case} \label{ng}

\begin{figure}
\makebox[\textwidth]{
\includegraphics[width=.5\columnwidth]{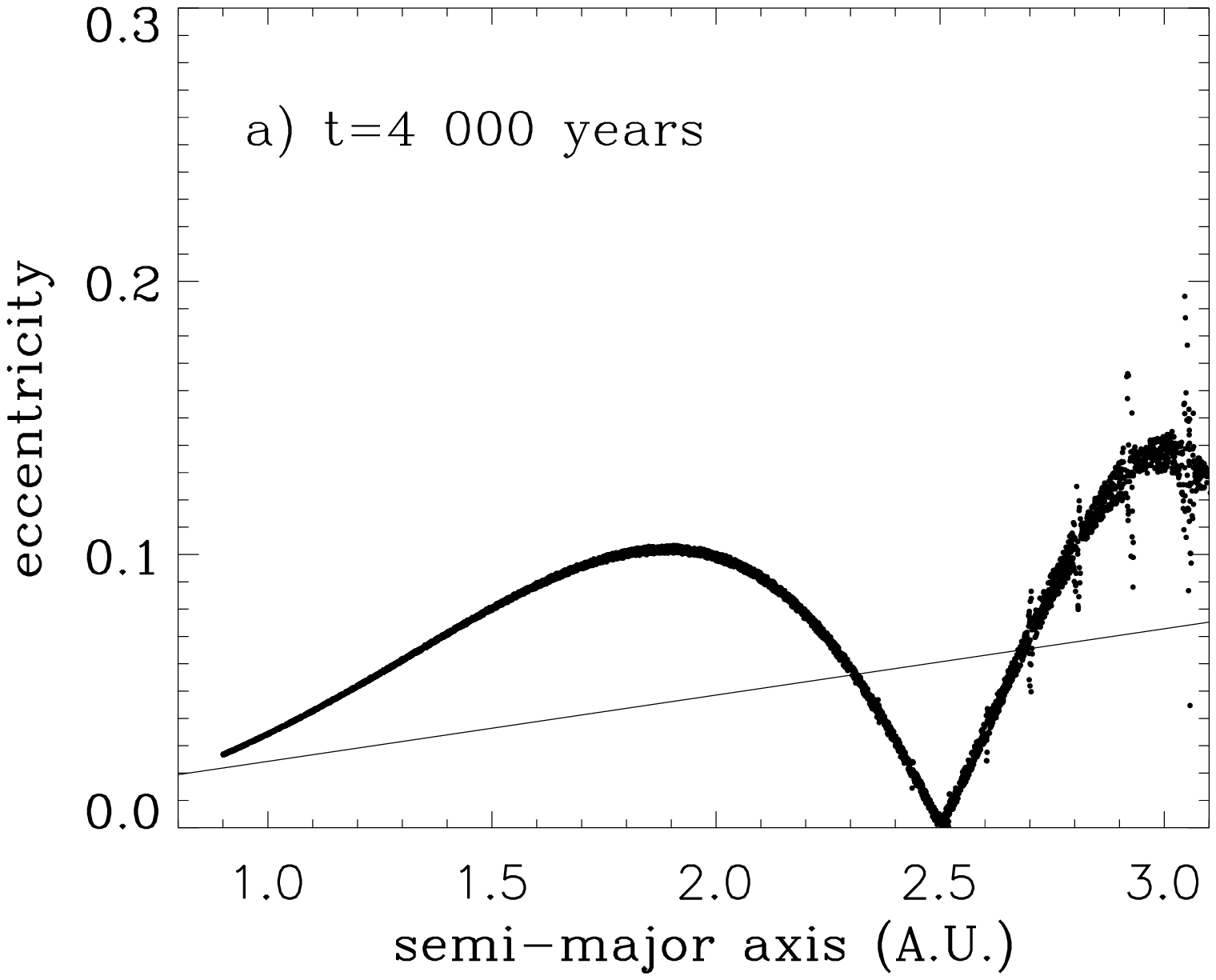}
\hfil
\includegraphics[width=.5\columnwidth]{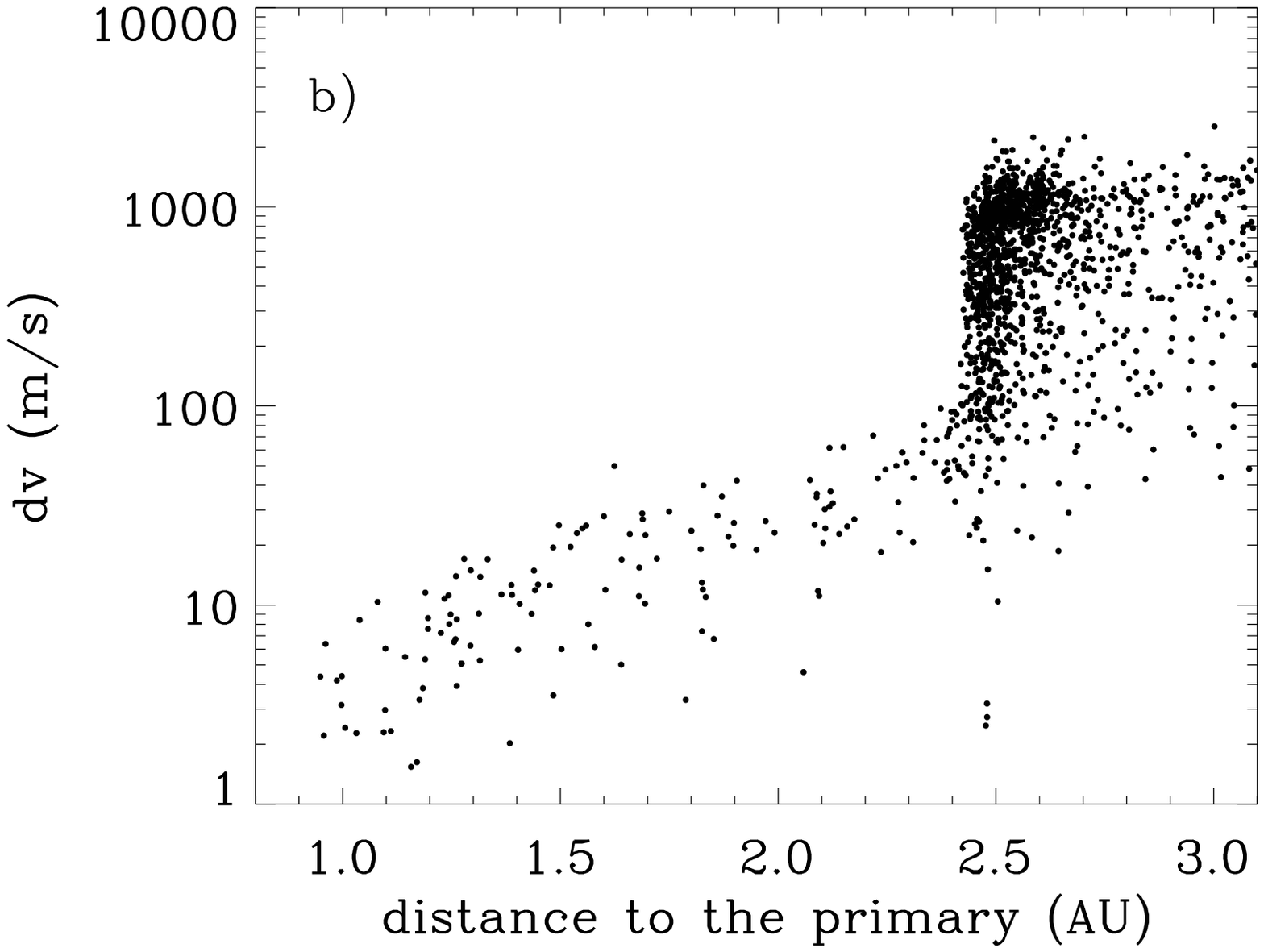}
}
\makebox[\textwidth]{
\includegraphics[width=.5\columnwidth]{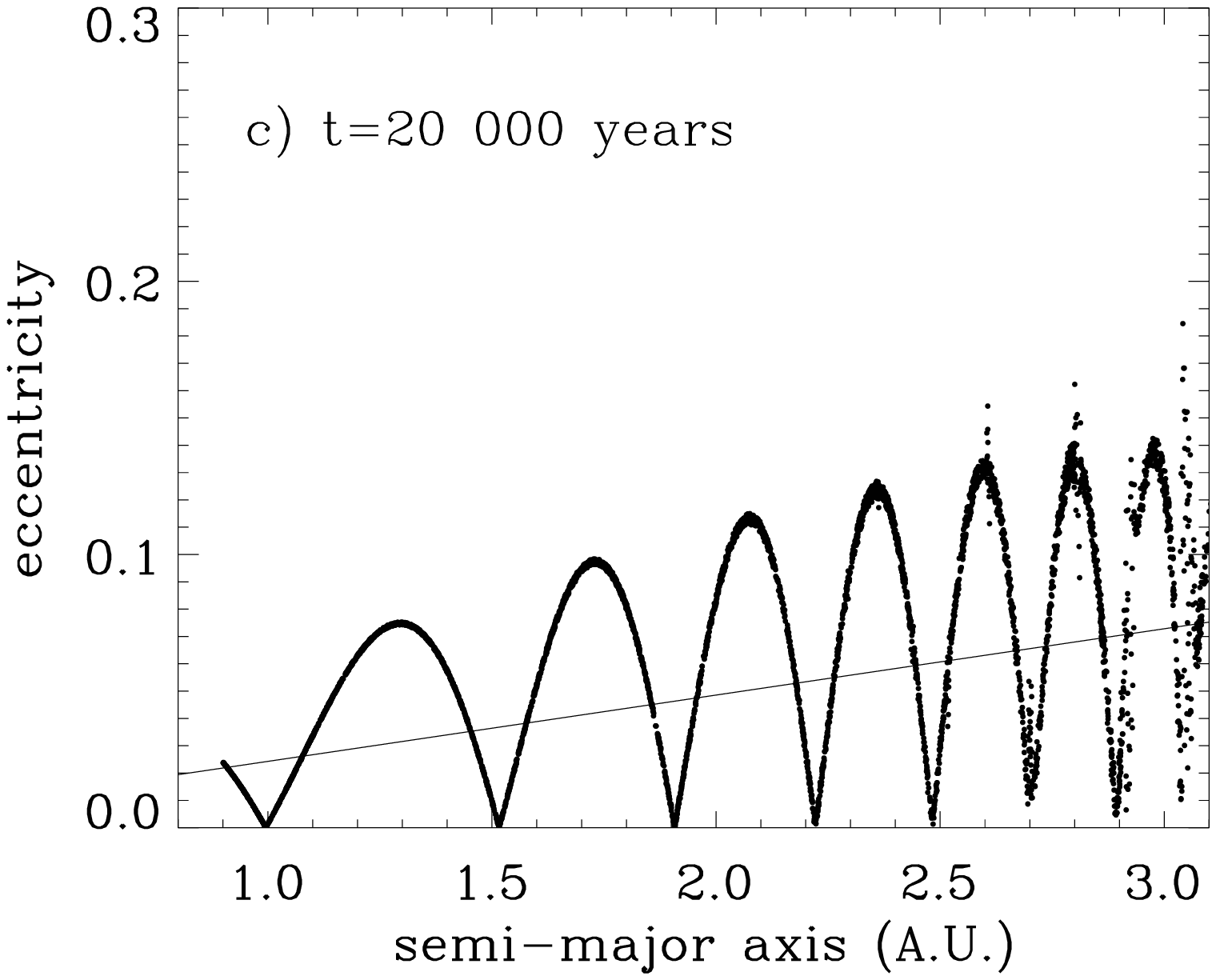}
\hfil

\includegraphics[width=.5\columnwidth]{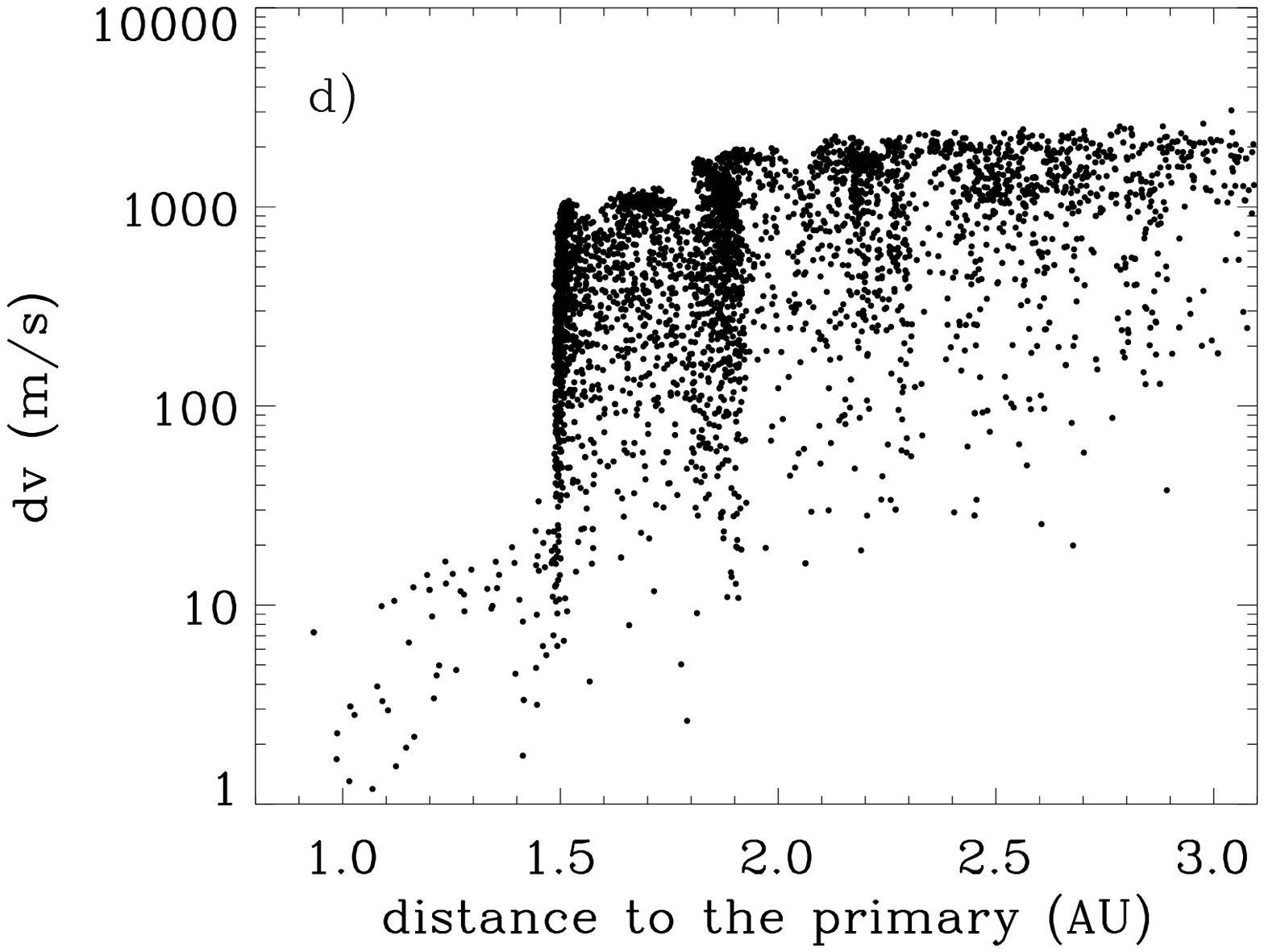}
}
\caption[]{Gas free case. Snapshots, after t$=4\times 10^{3}$ and t$=2\times 10^{4}$\,years, of the eccentricity distribution (left-hand side) and encounter velocity distribution (right-hand side) within the planetesimal population, as a function of semi-major axis (for the eccentricity) and radial distance to the primary ($dv$ panels). The straight line on the (e,a) panels indicates the value of the forced secular eccentricity $e_f$. The velocity distribution is obtained by recording all impacts in the $t\pm100\,$years time interval. The inward propagation of the high-$dv$ "wave", due to orbital crossing of neighbouring orbits, is clearly visible.
}
\label{nogas}
\end{figure}

In order to clearly identify the different mechanisms at play, we first present a fiducial and pedagogical run with no gas. Note that planetesimal physical sizes are irrelevant for this purely gravitational case. Figs.\ref{nogas}a and c show the classical build up of large secular eccentricity oscillations around the forced eccentricity $e_f$. These oscillations are due to the fact that, at each radial distance from the primary, eccentricities vary with time following a sinusoidal function 
\begin{equation}
e(a,t) = 2\,e_f\,\mid sin(ut/2)\mid = \frac{5}{2}\frac{a}{a_b}\frac{e_b}{1-e_{b}^{2}} \mid sin(ut/2)\mid 
\label{efor}
\end{equation}
whose frequency $u$ vary with semi-major axis \citep[e.g.][]{theb06}.
As has been pointed out in previous studies, such oscillations do not immediatly lead to high impact velocities, because neighbouring orbits are strongly phased. However, because the $a$ dependency of particle eccentricities increases with time (since $u$ depends on $a$), the $e(a)$-oscillations get narrower with time and orbits within one oscillation "wave" eventually cross, at which point very high $dv$ are suddenly reached \citep[see discussion in][]{theb06}. The radial location at which orbits cross moves inward with time following the empirical law derived by \citet{theb06}:
\begin{equation}
a_{cross} \sim 0.37
\frac{\left(1-e_{b}^{2}\right)^{1.07}}{e_{b}^{0.36}}
\left(\frac{M_{b}}{1M_{\odot}}\right)^{-0.39}
\left(\frac{a_{b}}{10\rm{AU}}\right)^{1.53}
\left(\frac{t}{10^{4}yr}\right)^{-0.36}
\,\rm{AU} \,.
\label{acr}
\end{equation}

Note that the location and timing of the orbital crossing does only weakly depend on the initial conditions for planetesimal orbits. We have chosen here the simplest and probably less unlikely case of initial circular orbits \citep[for more on this issue, see the discussion in][]{theb06}, but taking initial eccentric orbits will roughly lead to the same behaviour for $a_{cross}$, the only change being an additional free component to the encounter velocities.

From Fig.\ref{nogas}b, we see that $a_{cross}$, i.e. the location beyond which there is an abrupt increase of impact velocities, reaches $2.6$\,AU (the present location of the planet) in less than $4\times 10^{3}\,$years. At this point, encounter velocities increase by more than a factor $\sim 20$. However, even the short timespan before orbital crossing is not fully calm, because of sporadic high-$dv$ impacts (100 to 200m.s$^{-1}$) with particles in the nearby mean motion resonances clearly seen on Fig.\ref{nogas}a. In this gas-free case, the $r\sim 2.6\,$AU region is thus very hostile to low-$dv$ planetesimal accretion. 

The situation is less desperate closer to the primary. The regions inside 1.5\,AU is for instance protected from orbital crossing, and high-$dv$, for more than $2\times 10^{4}\,$years (Fig.\ref{nogas}d). This should in principle leave enough time for runaway accretion to produce embryos from kilometre-sized planetesimals. However, we shall see that gas drag completely obliterates these optimistic conclusions.

\subsection{Nominal Run with Gas} \label{gnom}

\begin{figure}
\makebox[\textwidth]{
\includegraphics[width=.5\columnwidth]{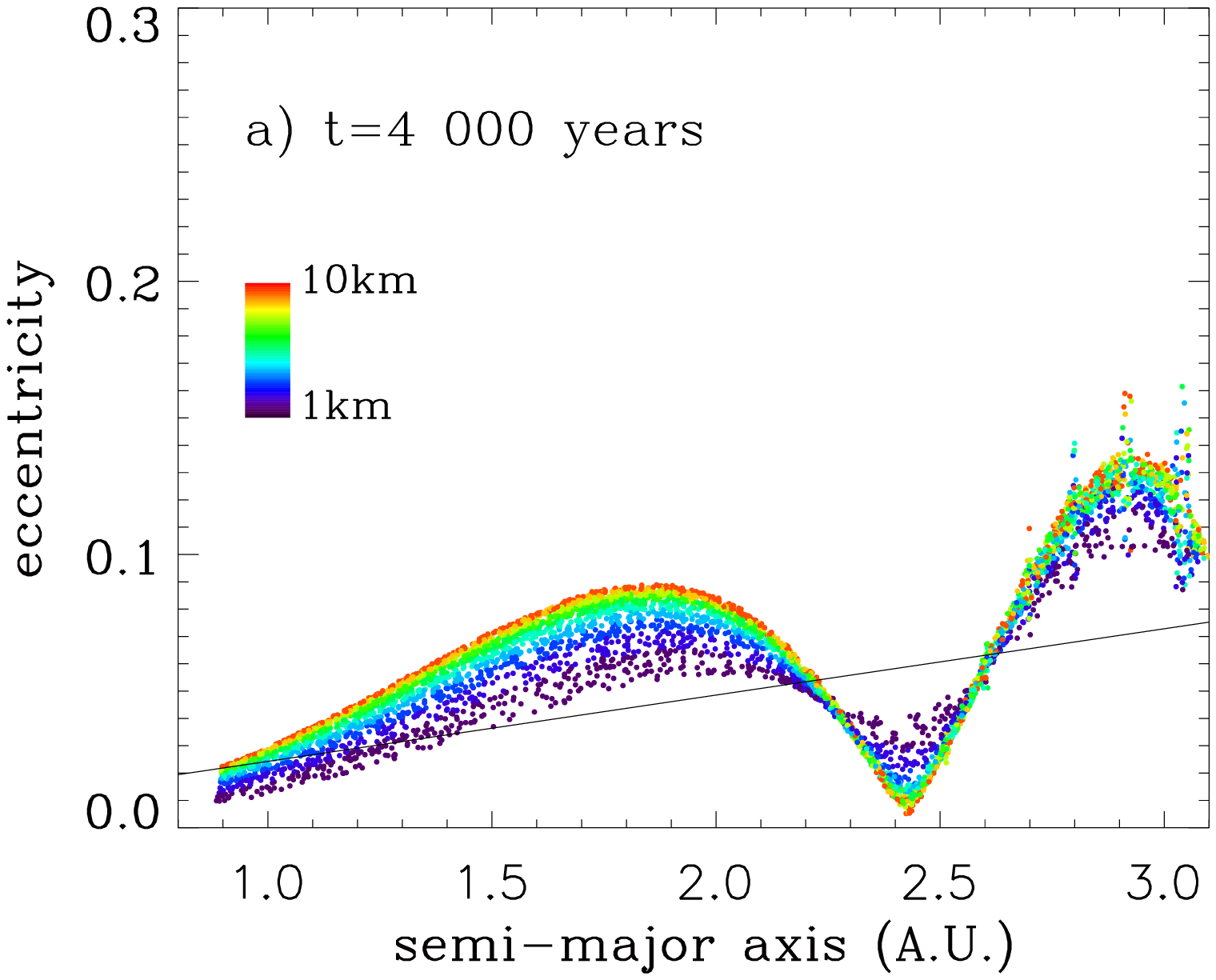}
\hfil
\includegraphics[width=.5\columnwidth]{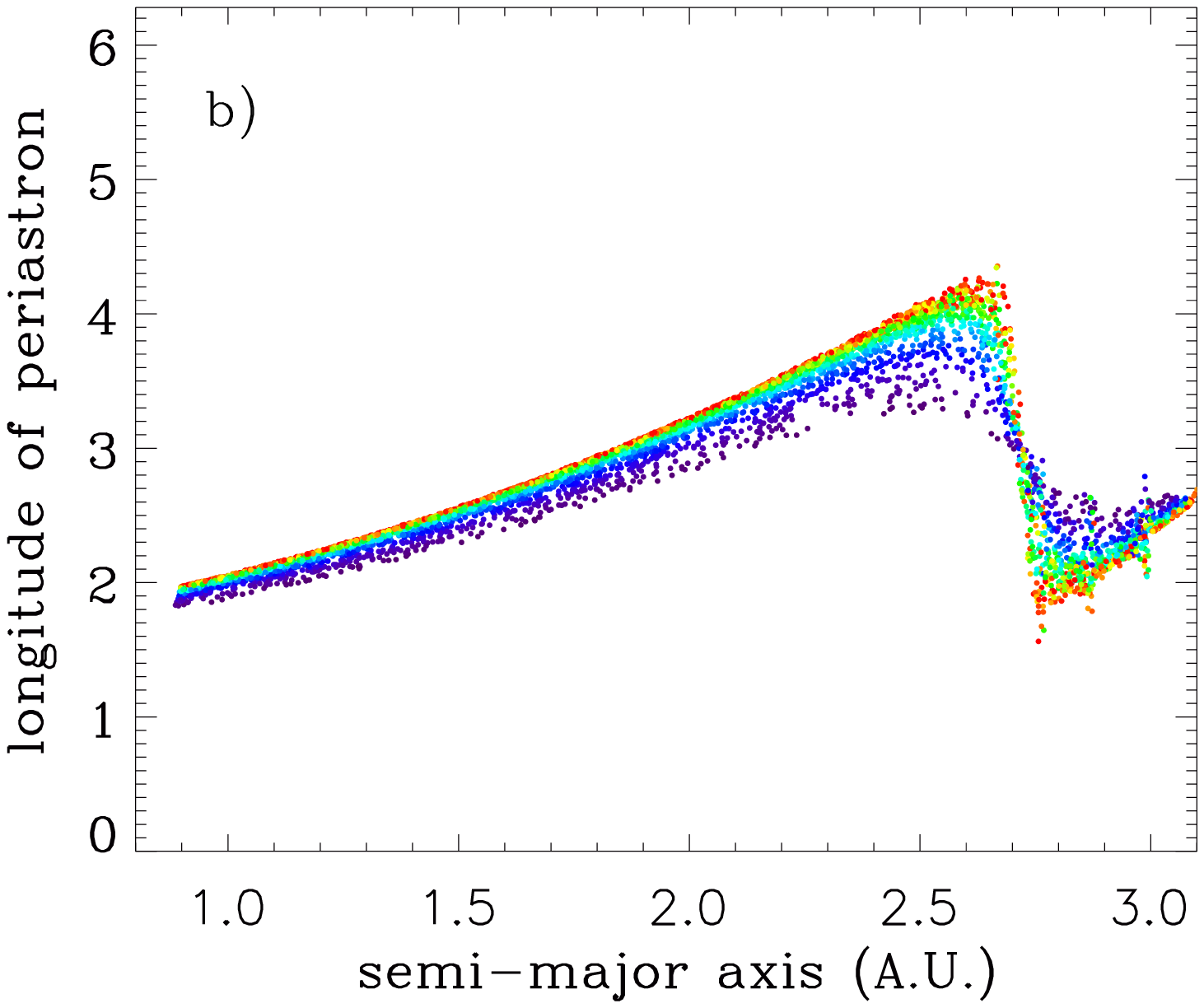}
}
\makebox[\textwidth]{
\includegraphics[width=.5\columnwidth]{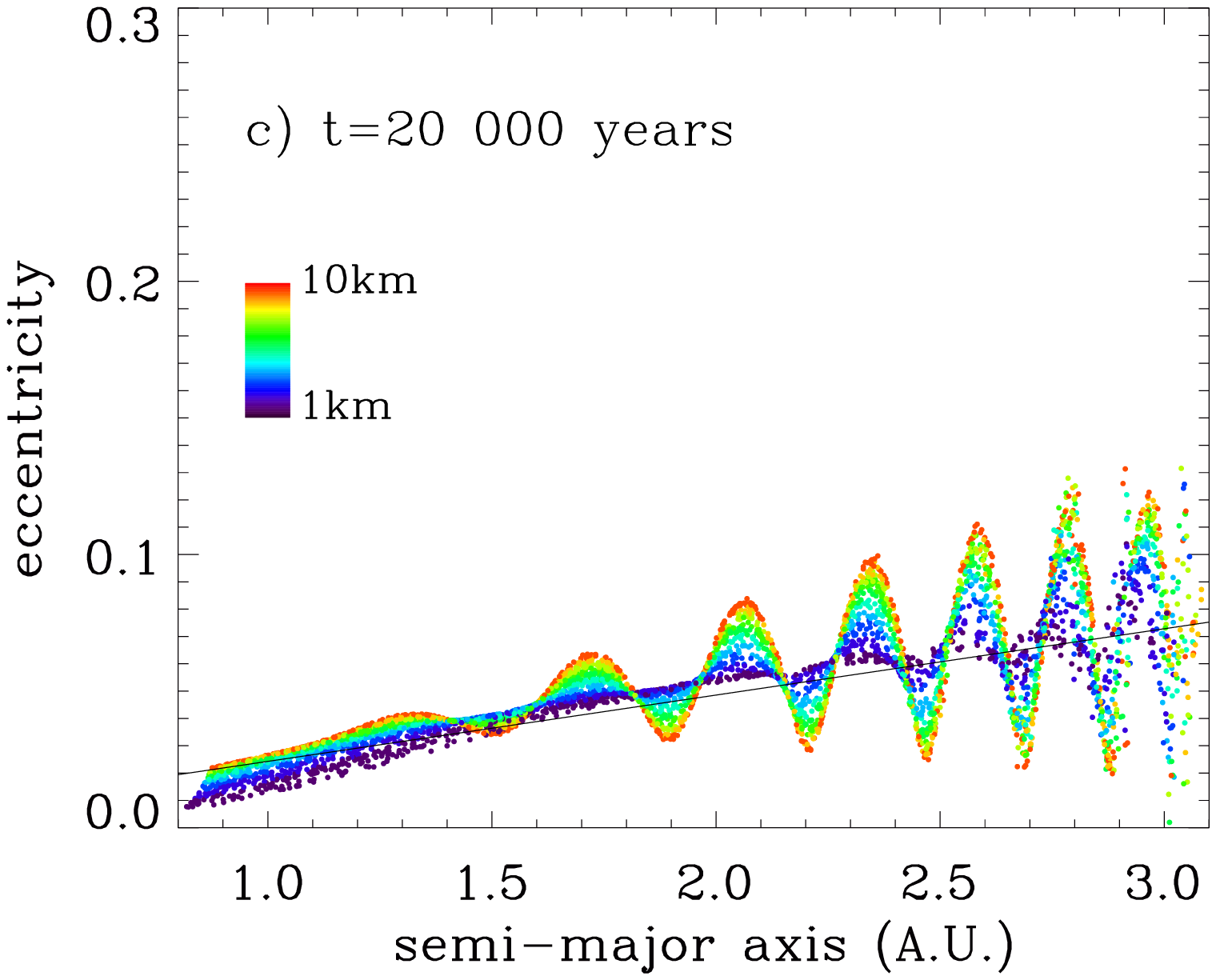}
\hfil
\includegraphics[width=.5\columnwidth]{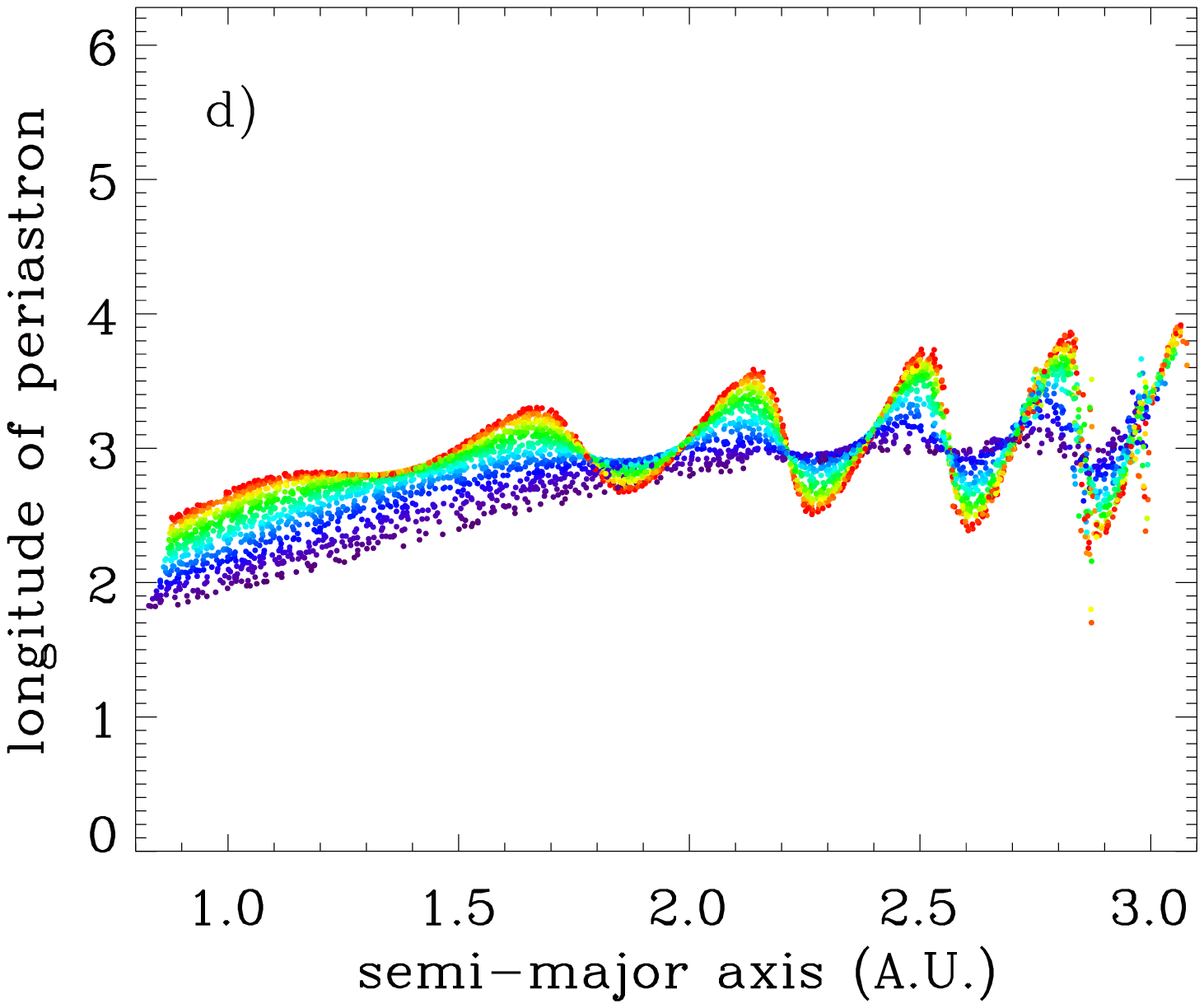}
}
\caption[]{Nominal case with gas friction from a 1xMMSN disc: eccentricity (left panels) and longitude of periastron (right panels) as a function of semi-major axis (the binary's longitude of periastron is 0), at $t=4\times 10^{3}\,$ and $t=2\times 10^{4}\,$years, for a population of planetesimals in the $1\leq s \leq 10\,$km size range.
}
\label{gasnom}
\end{figure}

\begin{figure}
\makebox[\textwidth]{
\includegraphics[width=.5\columnwidth]{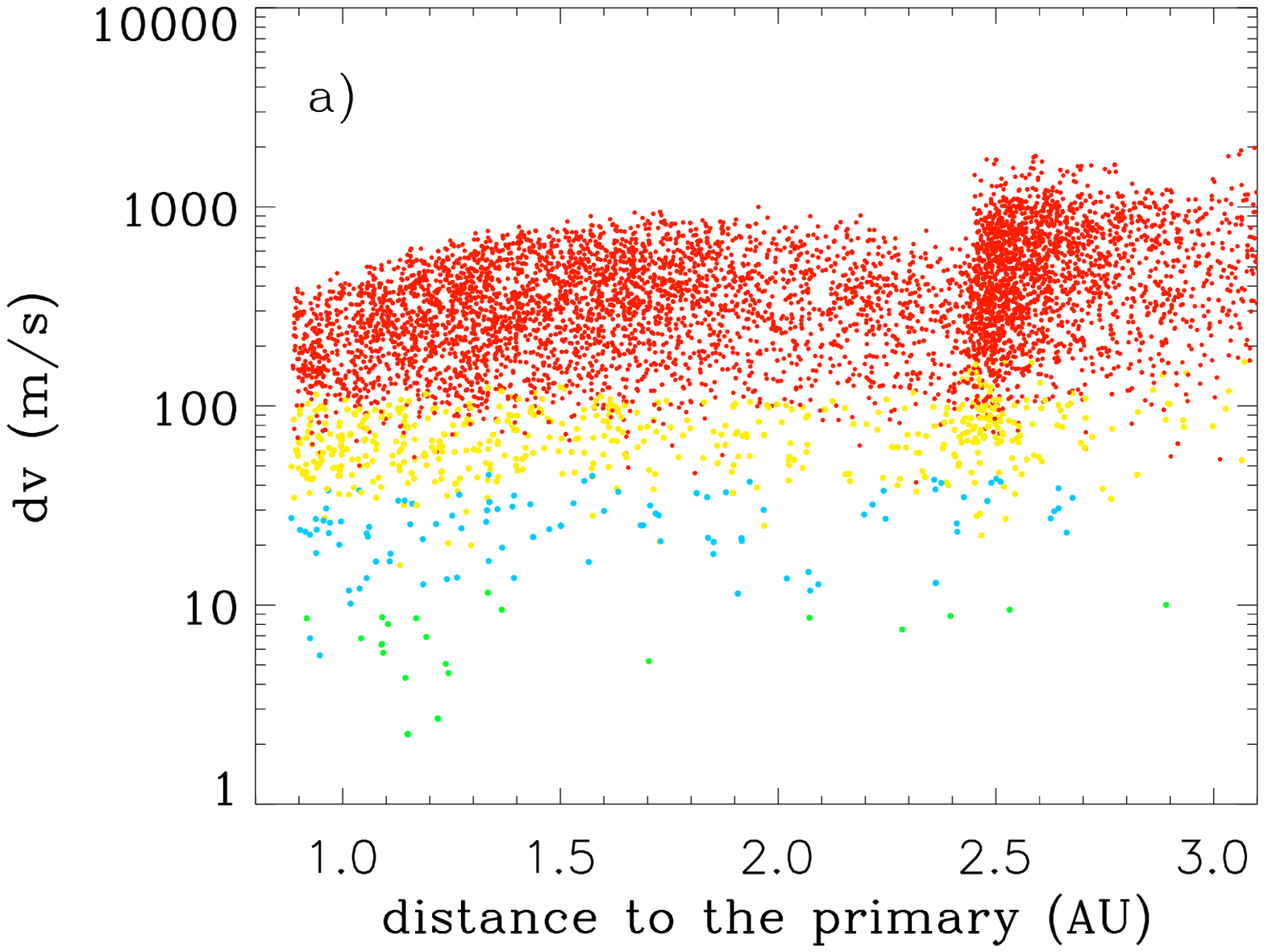}
\hfil
\includegraphics[width=.5\columnwidth]{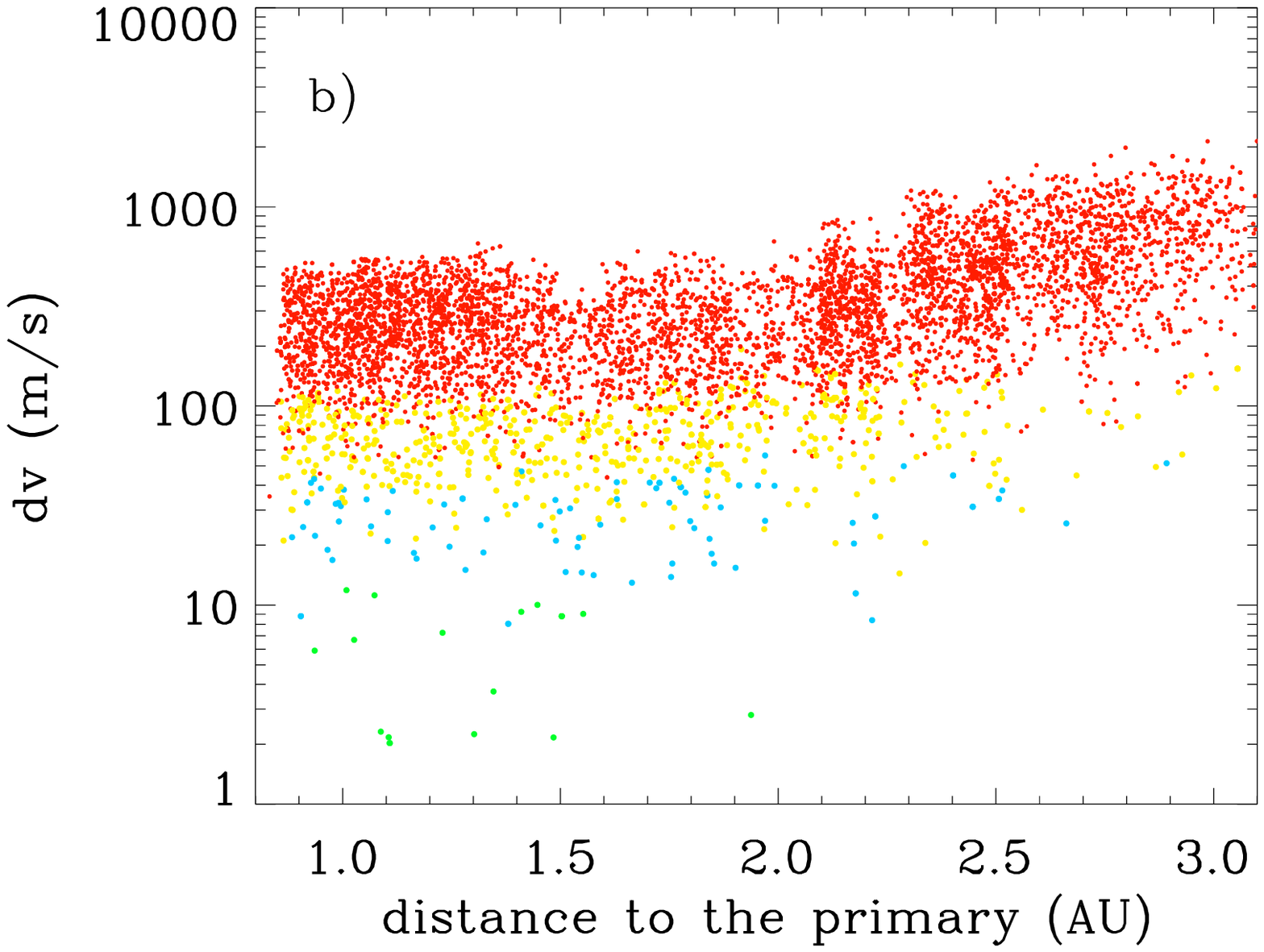}
}
\caption[]{Same nominal gas-friction run as in Fig.\ref{gasnom}: impact velocity distribution as a function of radial distance to the primary. The colours indicate the expected collision-outcome regime for each impact: "unperturbed" runaway accretion ($green$), perturbed accretion ($blue$), erosion ($red$). The yellow area is for impacts where $v_{ero1}\leq dv \leq v_{ero2}$, for which the collisional outcome is uncertain (see text for details).
}
\label{gasnomv}
\end{figure}

Gas drag radically changes the dynamical evolution of the system. Its main effect is to phase both planetesimal eccentricities (Fig.\ref{gasnom}a and c) and longitude of periastron $\omega$ (Fig.\ref{gasnom}b and d) according to their size $s$. 
As shown in Fig.\ref{gasnom}c, the initially large eccentricity oscillations are progressively damped. The system will eventually tend towards a steady state, where all eccentricities reach an equilibrium value $e_{(a,s)}$ depending on semi-major axis and size. In the innermost regions with higher gas densities, this steady state is reached relatively quickly for all particles in our 1-10\,km size range. Beyond $\sim 1.4\,$AU, however, the steady state has not been reached, at $t=2\times 10^{4}\,$years, for the biggest planetesimals. And at the location of the planet, 2.6\,AU, even the smallest 1\,km objects still have residual eccentricity variations at the end of the run (Fig.\ref{gasnom}c).

\begin{figure}
\makebox[\textwidth]{
\includegraphics[width=.5\columnwidth]{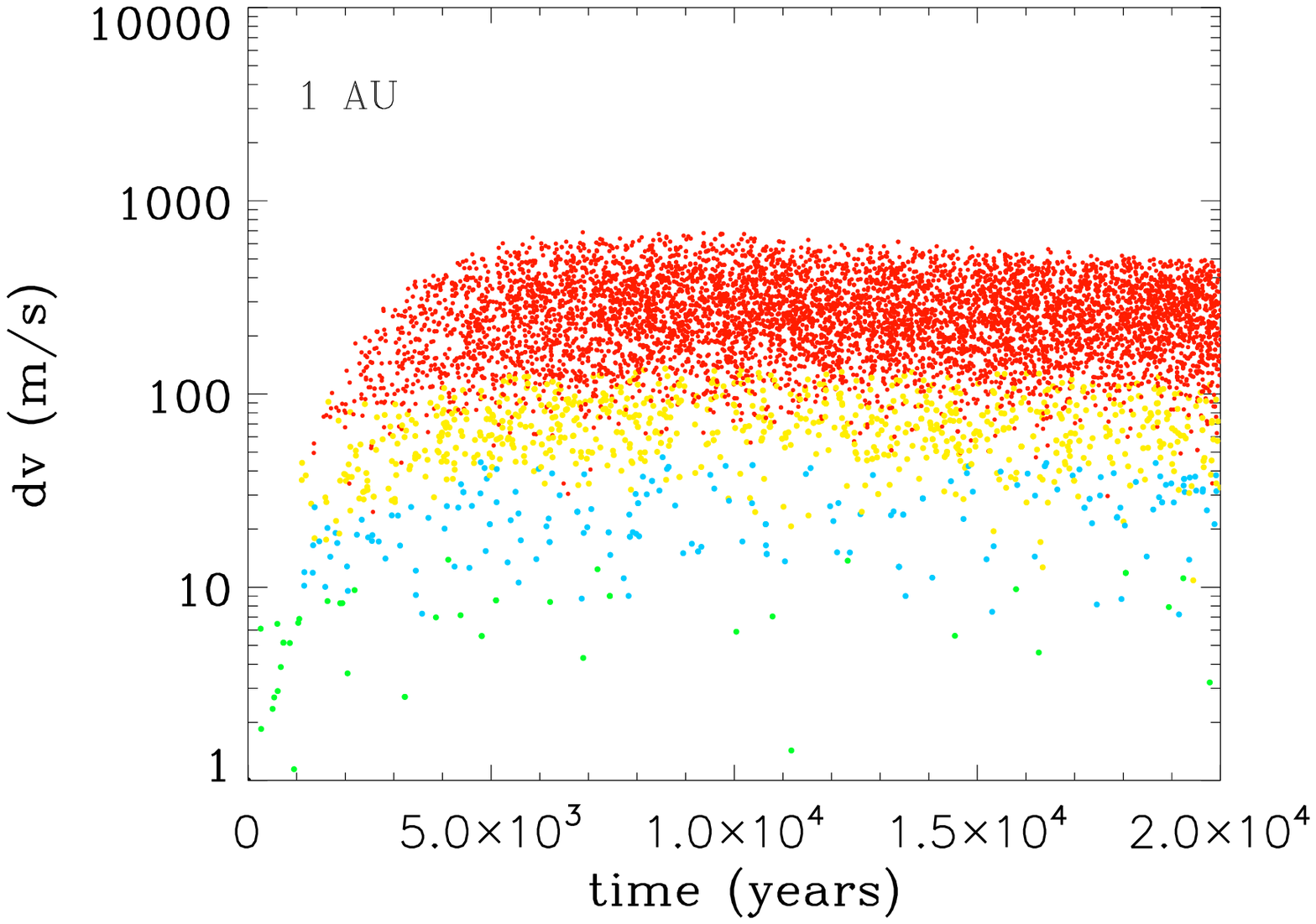}
\hfil
\includegraphics[width=.5\columnwidth]{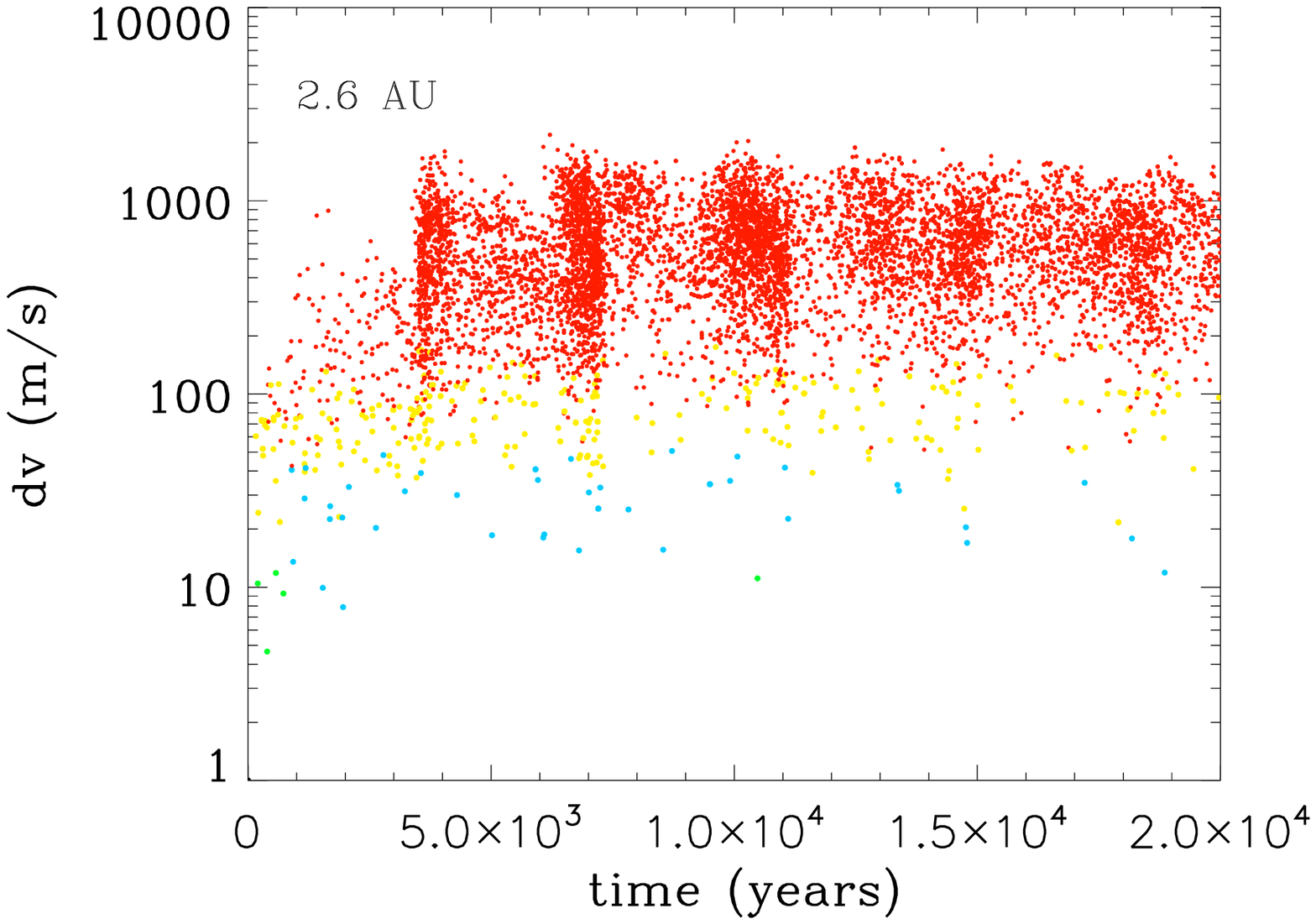}
}
\caption[]{Nominal gas drag case. Time evolution of the impact velocity distribution at 2 different locations in the disc. All impacts are recorded in a narrow ring of width $0.02\,$AU around the central location. The colour scale is the same as in Fig.\ref{gasnomv}. 
}
\label{gastime}
\end{figure}

These dynamical behaviours have clear consequences on impact velocities, which reach high values everywhere in the disc (Fig.\ref{gasnomv}). These high $dv$ are reached very quickly, a few 1000\,years, and this for two different reasons depending on location within the disc:
\begin{itemize}
\item In the inner disc, shortwards of $\sim$ 2-2.5\,AU, where gas densities are high, the differential phasing induced by gas drag, and the velocity increase that comes with it, is felt very early on, long before orbits reach a steady state with fixed $e$ and $\omega$. This is illustrated in Fig.\ref{gasnomv}a showing that, at $t=4\times 10^{3}$years, when no region of the disc has reached a steady state yet, $dv$ already reach values $\geq 300\,$m.s$^{-1}$ in the whole $\leq2.5\,$AU region.
\item In the $r \geq 2.5\,$AU region, the gas drag-induced $dv$ increase is weaker, but velocities nevertheless reach even higher values, $\sim 1000$m.s$^{-1}$, because of the purely dynamical orbital crossing effect identified in the gas-free runs. This is clearly illustrated by the $dv$ "jump" at $\sim 2.5\,$AU in  Fig.\ref{gasnomv}a.
\end{itemize}.
The velocity distribution at $t=2\times 10^{4}\,$years is remarkably similar to the one at $4\times 10^{3}\,$\,years. The only difference is in the outermost regions, where no $dv$ jump is longer visible, because gas drag is now the dominant $dv$-inducing mechanism in the whole system \footnote{This $t \sim 2\times 10^{4}\,$years time is thus the characteristic timescale for gas drag to dominate the whole system's dynamics}  (even if a steady state has $not$ been reached in these outer regions).
To illustrate these behaviours more clearly, we display in Fig.\ref{gastime} the temporal evolution of $dv$ in two opposite regions of the disc. At 1\,AU, differential phasing induces a high-$dv$ regime after $\sim 2\times 10^{3}$years, while a steady state is reached after $\sim 5\times 10^{3}$years. At 2.6\,AU (the planet location), three succesive phases can be distingued: a first phase, starting after only a few 100\,years, of moderate-to-high-$dv$ induced by sporadic impacts with objects in neighbouring resonances, followed by a second stage, at $t=4\times 10^{3}$years, when orbital crossing occurs and increases $dv$ to even higher values, and finally a third stage, starting around $1.5\times 10^{3}$years, when gas drag phasing progressively takes over as the dominant $dv$-inducing mechanism. 
Despite these differences, however, the important result is that a high-$dv$ regime, regardless of its different causes depending on location in the disc, is reached after only a few 1000\,years in both the inner and outer regions of the disc.

To estimate the global consequences of this high-$dv$ regime on the accretional evolution of a "real" population of planetesimals, one has to consider a parameter that has been ignored so far: the planetesimal {\it size distribution}. In our simulations, a flat distribution between 1 and 10\,km has been considered for the sake of simplicity, but real distributions should be more complex. This issue of the initial planetesimal size distribution is a difficult one. In models of planetesimal accretion in our solar system, a single "initial" planetesimal size is usually assumed \citep{maki98,koku00}, but this assumption is taken for the sake of simplicity and there should be some size dispersion in any realistic initial planetesimal population. The exact profile of this initial distribution is difficult to constrain. Firstly because it depends on the way planetesimals are formed from smaller grains and pebbles, a process which is still far from being fully understood, even if significant progresses have been made in recent years \citep{joha07,cuzz08}. Furthermore, the very concept of an "initial" size distribution can be questioned, as there might be a wide spread in the times at which km-sized objects appear in a given region of the disc \citep{cham10,xie10b}. These issues go well beyond the scope of the present paper, and we shall consider, following \citet{theb08,theb09} and \citet{xie09}, a Maxwellian distribution centered on $s=5\,$km. Such a relatively peaked disitribution is, in line with our conservative approach, a priori more accretion-friendly since it minimizes the rate of encounters between differently-sized objects. It also agrees with most planetesimal-formation scenarios' conlusion that there should be a privileged size for initial planetesimals. In practice, we weight each $s_1$--$s_2$ impact obtained in our run with a flat size distribution by a factor $f_{(s_1,s_2)}$ accounting for the Maxwellian distribution \footnote{The reason why we do not run a simulation with a Maxwellian distribution to start with is because, with a flat 1-10km distribution, we can get a good statistics on all possible impacting sizes $s_1-s_2$}.

\begin{figure}
\includegraphics[width=0.5\columnwidth]{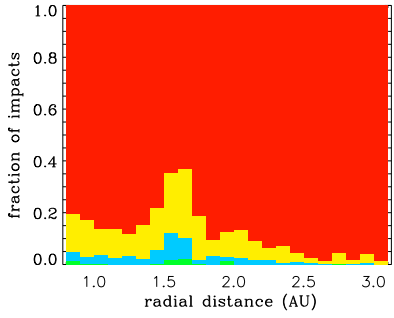}
\caption[]{Nominal gas drag case. Relative importances, at $t=10^{4}$years, of the 4 possible collision outcomes as a function of radial distance to the primary (the colour scale is the same as in Figs.\ref{gasnomv} and in Fig.\ref{gastime}). A Maxwellian size distribution centered on 5\,km is assumed.
}
\label{gasacc}
\end{figure}

The accretion/erosion behaviour of the whole disc is displayed, at $t=10^{4}$years, in Fig.\ref{gasacc}. As can be clearly seen, the rate of impacts leading to mass erosion or fragmentation is $\geq 80$\% everywhere, except around 1.6\,AU where it is $\sim65$\%. When discarding the impacts with "uncertain" (yellow) outcome, the level of accreting impacts is less than 10\% everywhere. Moreover, among these accreting impacts, most of them are in the "perturbed" mode (blue), with almost no impact allowing runaway growth (green). We have tried different size-distributions and always found the same global accretion-hostile trend, except for extremely peaked, and probably unrealistic, distributions. Only with much larger planetesimals can this negative trend be reversed (see Sec.\ref{lplan}).

\subsection{Inclined Binary} \label{incli}

\begin{figure}
\includegraphics[width=0.5\columnwidth]{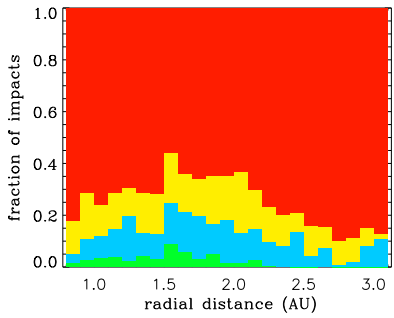}
\caption[]{Same as Fig.\ref{gasacc}, but with a binary inclined by $2^{o}$ with respect to the circumprimary disc.
}
\label{incli2}
\end{figure}

\citet{xie09} have shown that a small inclination between the binary and the circumprimary disc can help accretion by segregating particle inclinations according to sizes, thus favouring impacts between equal-sized bodies. We explored this possibility for the present HD196885 case, assuming a small inclination of $2^{o}$ for the binary. As shown in Fig.\ref{incli2}, an improvement is obtained compared to the coplanar case (Fig.\ref{gasacc}). Accreting impacts now make up between 5 and 20\% of all impacts in most of the disc. However, the system is still globally hostile to accretion everywhere (between 60 and 80\% of "red" impacts), especially at the location of the planet (2.6\,AU) where the fraction of eroding impacts is in excess of 90\%. 
We have tried several different values for the binary inclination in the $1^{o}\leq i \leq 10^{o}$ range, and always end up with results very similar to those displayed in Fig.\ref{incli2},i.e., a system that is globally very hostile to accretion.

Strengthening our conclusions is also the fact that the vertical segregation according to particle sizes obtained by \citet{xie09} is probably unrealistically high. This is because it has been obtained for an axisymetric gas disc, whereas a real gas disc would get eccentric and tend to diminish the size-sorting effect. Our results for an inclined binary, also obtained with an axisymetric gas disc, should thus be considered as a best-case scenario regarding planetesimal accretion; a best-case scenario for which no accretion is possible in the whole $\geq0.9\,$AU region.

\subsection{Large planetesimals} \label{lplan}

\begin{figure}
\makebox[\textwidth]{
\includegraphics[width=.5\columnwidth]{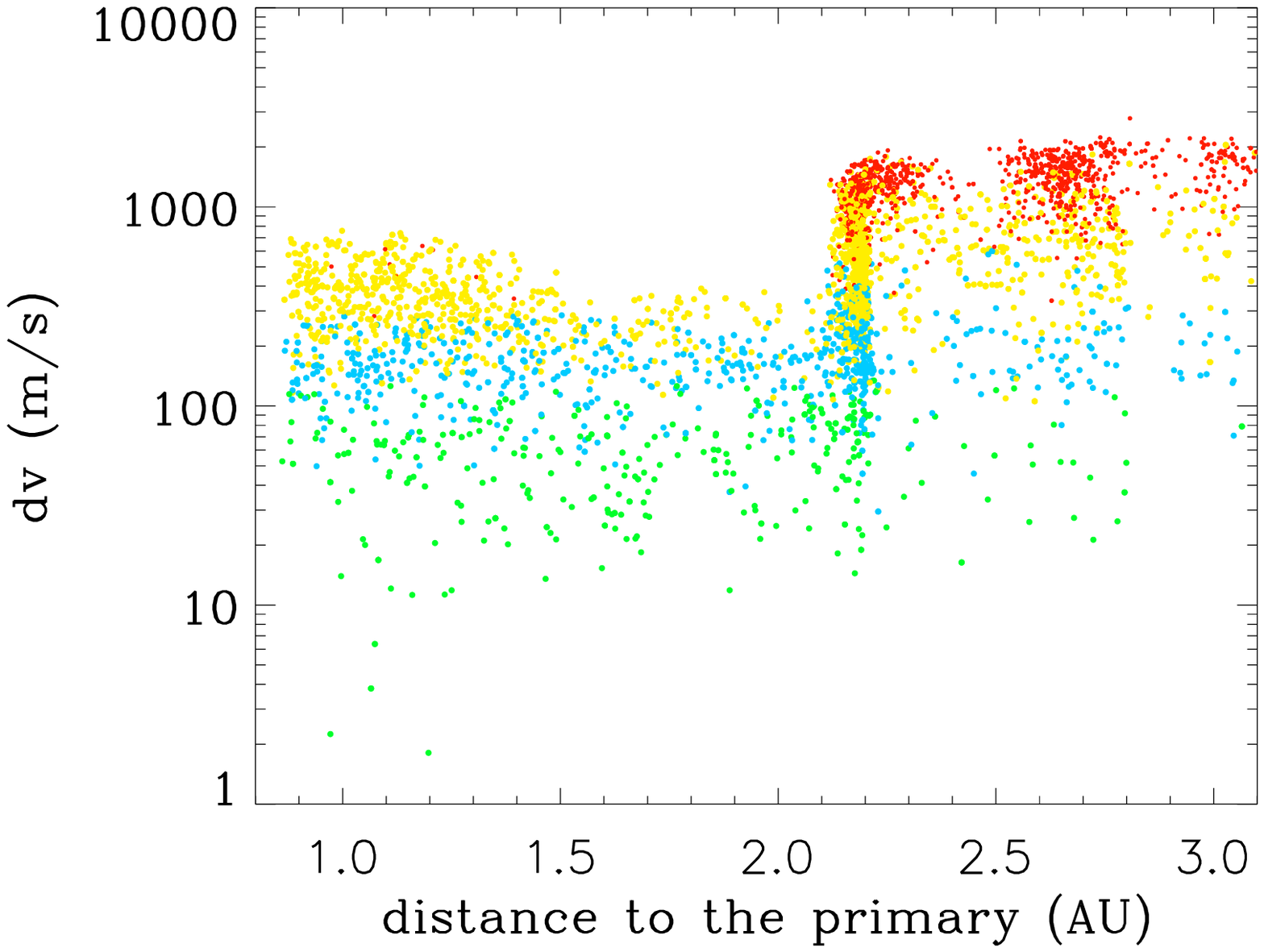}
\hfil
\includegraphics[width=.5\columnwidth]{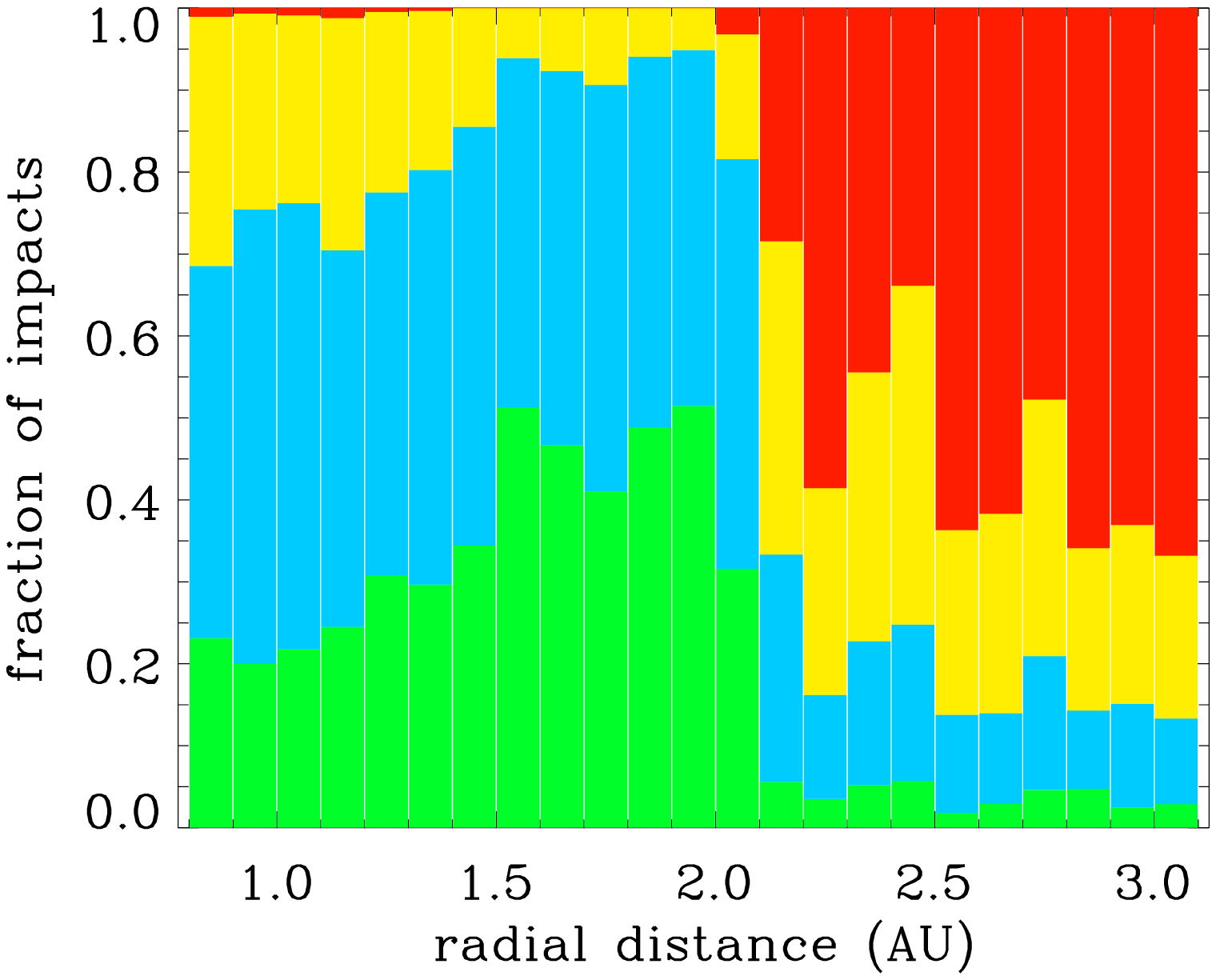}
}
\caption[]{Large planetesimals ($10\leq s \leq 100\,$km) case. Encounter velocity distribution at $t=10^{4}\,$years (left) and respective balance between collision outcomes (right). The colour scale is the same as in Fig.\ref{gasacc}. 
}
\label{big}
\end{figure}

As suggested by \citet{theb08} and, using slighly different assumptions, by \citet{beau10}, another possible solution to the accretion-hostile-velocities dilemma is to start from larger planetesimals. We explore this hypothesis by considering a population of initial planetesimals in the 10-100\,km range. Taking larger objects reduces the impact of gas drag and its subsequent effect on increasing $dv$ among planetesimals. This is what is observed in the inner regions of the disc in Fig.\ref{big}a, where impact velocities are lower than for the 1-10\,km population (Fig.\ref{gasnomv}a). This situation is more favourable for accretion, a tendency that is amplified by the fact that, for a given $dv$, bigger planetesimals are more resistant to impacts than smaller ones (in the gravity regime valid for objects in the $\geq$ 1-10\,km range). As a result, at $t \sim 10^{4}\,$years, the whole region shortwards of $\sim 2\,$AU is accretion-friendly for 100km planetesimals (Fig.\ref{big}b). In the outer regions, however, the situation is radically different: the values of $dv$ are much higher because of the secular orbital crossing effect. In these outer gas-poor regions, large planetesimals behave almost as test particles in the gas-free case. In regions of orbital crossing, impact velocities are high enough to lead to erosion even for 100\,km objects (Fig.\ref{big}b). As in the gas-free case, the high-$dv$ orbital crossing "front" reaches 2\,AU in $\sim 10^{4}\,$years, but the 2.6\,AU location is reached in less than $4\times 10^{3}\,$years.

\section{Discussion} \label{discu}

The previous results all tend to indicate that the HD196885 circumprimary disc is strongly hostile to the accretion of kilometre-sized bodies. This is especially true of the region at 2.6\,AU from the primary, i.e., around the current location of the detected planet. Planetesimals are basically caught between a rock and a hard place: in the inner regions secular perturbations combined to gas drag induce high-$dv$ between all non-equaly-sized objects, while in the outer regions high-$dv$ are due to secular perturbations $alone$, which make orbits cross within a few thousand years. 

\subsection{Limitations, robustness of our conlusions}

As underlined earlier, our numerical exploration relies on several simplifications. It has also many free parameters that cannot be all thoroughly explored. How robust are our conclusions, especially regarding the 2.6\,AU region, with respect to these limitations? 

\paragraph{Timescale.}
We first note that timescale is not a critical issue: high-$dv$ are reached almost everywhere after only a few $10^{3}$years. The only change with time is that gas drag progressively takes over as the dominant $dv$-inducing process even in the outer regions. 

\paragraph{Binary inclination.} 
As shown by \citet{xie09}, a small inclination between circumprimary disc and binary orbital plane acts in favour of planetesimal accretion. This point is all the more appealing because this inclination is in most cases an unconstrained parameter, even for close binaries, for which the assumed "coplanarity" is not constrained to less than $i_b \sim 10^{o}$ \citep{hale94}. In the present case, however, even if the situation improves significantly for $i_b$ of a few degrees, it is far from enough to reverse the general accretion-hostile trend of the system (Fig.\ref{incli2}). In addition, as pointed out in Sec.\ref{incli}, the vertical size-sorting of planetesimals, and its effect in favour of accretion, is probably overestimated in a circular gas-disc case. As a consequence, we believe our conclusions of an accretion-hostile disc with a slighly inclined binary to be relatively robust.

Of course, even higher values of $i_b$ could be possible, as might be suggested by the preliminary long-term stability study of \citet{chau10}. These high inclination cases cannot be explored with the present model, and will be the purpose of a forthcoming general study devoted to this issue (Xie et al., submitted), but it seems very unlikely that impact velocities should be reduced in such high-$i_b$ systems, especially when the Kozai regime sets in.

\paragraph{Gas disc profile.}
We have run several additional simulations (not shown here), exploring different gas disc profiles and densities. In almost all cases, results are roughly comparable to the nominal case: high, and accretion-inhibiting impact velocities in the whole $r\geq 0.9\,$AU region. Only for very tenuous discs do we get an accretion-friendly inner region for more than $10^{4}$years, which is basically the gas-free result displayed in Fig.\ref{nogas}. It coud be argued that this gas-free stage is the one towards which the real circumprimary disc is naturally evolving, since primordial gas is expected to be removed after a few $10^{6}\,$years in protoplanetary discs, or even less than that for close binaries \citep{ciez09}. In fact, \citet{xie08} have shown that some orbital rephasing could occur during the gas removal phase, possibly rending the disc accretion-friendly again. However, as shown by \citet{theb08}, it is unlikely that planetesimals could survive the long accretion-hostile period $before$ gas removal without being grounded to dust and removed by gas friction. The only solution would be that planetesimals form late in the disc's history, when few or not gas is left, but this hypothesis conflicts with all planetesimal formation models. In any case, let us stress that even in this unlikely scenario, the region around $2.6\,$AU would still be hostile to planetesimal accretion, due to secular effects $alone$.  

There is however another important gas-disc-related issue to consider here, i.e., that all our runs follow the same crude simplification, inherent to our approach, of a static axisymetric gas disc. Nevertheless, as discussed in Sec.\ref{model}, we expect planetesimals imbedded in evolving gas discs to have impact velocities that are even higher than in the fiducial static case \citep{paard08}. So here again, our $dv$ estimates do give a conservative lower estimate.

\paragraph{Gas self gravity.}
Another simplification of our model, as well as of most previous studies of planetesimals in binaries, is the neglect of the gas disc's gravity. To our knowledge, the only published studies taking into account disc gravity are \citet{kley07} and the very recent work by \citet{frag11}.
These pioneering studies present results with a limited number of particles (which is the price to pay for including the disc's gravity) that does not allow accurate $dv$ estimates, and for which it is difficult to untangle the effect of gravity from that of gas drag. However, it appears clearly that the global qualitative effect of disc gravity is to further {\it increase} impact velocities, by adding an additional jitter to the eccentricity and periastron evolution of planetesimal orbits \footnote{This is also the preliminary conclusion of \citet{marz08}, who also investigated disc gravity, but in the different context of a $circumbinary$ disc.}.
We do thus expect our gas drag-only simulations to here again give a lower limit for "real" impact velocities.

\subsection{A planet in an accretion-hostile environment?}

That most of the circumprimary disc is too excited to allow planetesimal accretion seems to be a relatively robust result for the HD196885 system. And yet there $is$ a planet well inside this accretion-hostile region. There are basically four potential solutions to this paradox: either 1) the planet could form in situ by being able to bypass the mutual-planetesimal-accretion phase, or 2) it was formed elsewhere and was later injected at its present location, or 3) the binary had a different orbit during its early history, or 4) the planet did not form by core-accretion but by direct disc instability.

\subsubsection{Bypassing the kilometre-sized planetesimals accretion phase?}

The most obvious potential way to bypass this stage is if planetesimals were formed big, i.e., not in the kilometre-sized range but rather in the 50-100\,km one. Interestingly enough, this big-initial-planetesimals scenario seems to be the one favoured by the most recent planetesimal-formation models of \citet{joha07} and \citet{cuzz08}. Moreover, according to \citet{morb09}, there seems to be observational evidence for an initial population of large, $\geq 100\,$km, bodies in the asteroid belt, even if this conclusion has been recently questionned, for different reasons, by \citet{mint10} or \citet{xie10b}. 
For the present problem, however, there are some issues with this large-planetesimals hypothesis. The first one is that, to our knowledge, there has been no study of how these big-planetesimal formation scenarios, for which several problems remain to be solved even for a normal single-star environment, could proceed in the highly perturbed environment of a close binary. The second, and more problematic issue for the specific HD196885 case is that, even $if$ initial planetesimals are $\sim 100\,$km big, the region around 2.6\,AU is still hostile to accretion (see the "large planetesimals" run in Fig.\ref{big}). We ran an additional test simulations and found that the 2.6\,AU region becomes accretion-friendly only for bodies with sizes $\geq 250\,$km. It is far from being assured that planetesimals can be born this big, especially in a highly perturbed close binary.

Another possibility to overcome the mutual-planetesimal-accretion hinder, even in the case of small planetesimals, is the so-called "snowball" growth mode first encountered in the simulations of \citet{paard10} and investigated in more detail by \citet{xie10b}. In this scenario, planetesimals grow preferentially by sweeping up of small dust particles, provided that the local mass density of solids contained in dust, $\rho_{S(dust)}$, exceeds that contained in planetesimals, $\rho_{S(plan.)}$. The snowball growth mode is especially appealing for dynamically excited systems such as close binaries because the accretion of dust onto planetesimals should tolerate much higher velocities than the mutual accretion of the planetesimals themselves. Exactly how much higher is not clear yet, as there exists to our knowledge no published study of the velocity dependence of the accretion efficiency of dust on large targets, at least for the $dv \geq 500$m.s$^{-1}$ regime encountered here \footnote{The "high velocities" dust impacts considered in laboratory experiments such as those of \citet{teis09} do not exceed 50-100m.s$^{-1}$}.
One additional issue is of course whether or not the $\rho_{S(dust)}\geq \rho_{S(plan.)}$ criteria is met in real systems. As shown by \citet{xie10b}, this condition could be fulfilled if there is a significant spread in the times at which planetesimals do appear in the system, as would for instance be expected for the \citet{cuzz08} scenario. But even if that is not the case, the $\rho_{S(dust)}\geq \rho_{S(plan.)}$ condition could be met later on in a perturbed system's evolution, when the quantity of dust produced by destructive planetesimal-on-planetesimal impacts exceeds the mass left in the remaining unshattered planetesimals \citep{paard10}. Both of these cases have so far been studied in preliminary works using simplified prescriptions: analytical expression for the growth of initially isolated planetesimals in \citet{xie10b}, and 2-D simulations for the \citet{paard10} studies of planetesimal re-accretion of impact-produced dust. They both will be quantitatively re-investigated in forthcoming studies (Paardekooper et al., 2011 and Xie et al., 2011, both in preparation).

\subsubsection{Embryo migration, Planet-Planet Scattering?}

Apart from the strongly revised versions of the planet-formation scenario presented above, there exist other potential solutions to the inhibition of the planetesimal-to-embryos stage. The first one is that embryos form in accretion-friendly regions closer to the primary and later migrate outward to the planet's present position, where they can continue to grow because the final embryo-to-planet stage is much less affected by binary perturbations (see Sec.\ref{intro}). This scenario has been quantitavely investigated by \citet{payn09}, who showed that a fraction of the embryos formed in the inner regions can indeed later move out. The relative amplitude $\delta a/a$ of this outward migration can reach 0.3 to 0.8. This is however not enough for the present case, because it means that the innermost possible origin for an embryo having moved to 2.6\,AU is $\sim 1.4\,$AU, i.e., a region that is still highly hostile to planetesimal accretion (see Fig.\ref{gasacc}). 

On a related note, one could imagine that the planet fully formed in the inner, accretion friendly regions and was later ejected by gravitational interactions with a second, yet undetected, planet. There is however a major problem with this scenario, which is that the accretion-friendly region around HD196885A is very narrow. By running an additional simulation focusing on the inner $r \leq 0.9$\,AU disc, we indeed find that the limit between the eroding (majority of "red" impacts) and accreting (majority of "green" $+$ "blue" impacts) regimes is located at around 0.4\,AU from the primary.
This would mean that 2 giant planets \footnote{the second undetected planet would also need to be massive in order to be able to perturb HD196885b} would have to form within 0.4\,AU from HD196885A. Such an hypothesis seems to be ruled out by all planet-formation scenarios.
Another counter argument is that a second planet closer to the primary would have been detected in radial velocity measurements (Chauvin, personal commuinication).

\subsubsection{Wider initial binary?}

Another possibility is that, during these early stages of its existence, the binary's orbit was different from what it is today. This might happen because most stars are expected to be born in clusters, which are initially compact, thus allowing frequent interactions between neighbouring stars. The simulations of \citet{malm07} have shown that, for the typical cluster they considered, binaries with moderate-to-large separations do suffer early encounters that have on average shrinked their orbit. Interestingly, HD196885 is just at the limit present-day separation, $\sim 20\,$AU, above which these orbit-shrinking effects are found to be significant \citep[see Fig.4 of][]{malm07}. There is thus a non-negligible chance that its initial orbit, during the planet formation phase, was wider than what it is today. How much wider would it need to be in order to allow planetesimal accretion at 2.6\,AU? We ran a series of test simulations with increasing values of the binary separation and found that the $r \sim 2.6\,$AU region becomes accretion-friendly for $a_b \geq 45$AU (for the same value of the eccentricity $e_b=0.42$). This means that the binary's orbit would have had to shrink by at least 24\,AU during its early history. Is such a value realistic? Unfortunately, it is impossible to judge from the \citet{malm07} simulations, since no information about the amplitude of orbital compaction was presented in this study (this important issue should clearly be investigated in future studies). However, such a large, $\sim$ 20-25\,AU, change in semi-major axis does a priori appear unlikely for a binary that is just at the limit separation for which orbital change become significant.

\subsubsection{Formation by disc instability?}

If none of the aforementioned solutions works, then a more radical alternative might be considered, i.e., to forgo the core-accretion scenario altogether. This is the conclusion recently reached by \citet{duch10}, who argued that the shorter disc lifetime in tight binaries makes it difficult to form giant gaseous planets. This study advocates a violent formation process, by direct disc fragmentation, for planets in $a_b \leq 100\,$AU binaries, a hypothesis that seems to be supported by the fact that exoplanets within $a_b \leq 100\,$AU binaries are significantly more massive than those within wide binaries or single stars. 
Such a scenario is usually invoked to explain the formation of planets at large, $\geq 100\,$AU semi-major axis \citep{bole09}, but \citet{duch10} argues that, in the specific context of tight binaries, it could also work for planets much closer to their star. His main argument is that circumprimary discs in close binaries could be more compact and denser than discs around single stars \footnote{\citet{duch10} argues that the lower (sub)millimetre fluxes measured for discs in $<100\,$AU binaries do not prevent these discs from being as massive (and thus denser) as single-star ones, because such truncated discs get optically thick to their own emission (see Fig.2 of \citet{duch10}).}, thus potentially favouring gravitational instability, and also because perturbations of the close stellar companion could give an additional trigger to instability \citep{boss06}.

However, this alternative explanation should be considered with some caution. Several studies have indeed shown that the instability scenario does also encounter major difficulties in the context of close binaries, and that no circumprimary disc gets dense enough to be unstable. As an example, \citet{nels00} or \citet{maye05} conclude that instability is severely hampered by the presence of a close ($\leq 50-60\,$AU) companion. 
This issue is thus far from being settled yet, and further, more detailed investigations are clearly needed to assess if disc-fragmentation can be considered as a viable alternative formation channel in close binaries.

\section{Summary and Conclusions} \label{conclu}

The planet of the HD196885 system is to this day the one that has the most perturbed orbit amongst all the $\sim 100$ exoplanets detected within binaries. Leaving aside the issue of the long term stability of its orbit, the question of how this planet could form under such extreme conditions is a critical one, that might have implications on our understanding of the planet-formation process in general.

We have investigated one specific leg of the planet-formation process, the intermediate stage leading from kilometre--sized planetesimals to protoplanetary embryos, as it is the one that is probably the most affected by perturbations in close binaries. We numerically follow the evolution of one crucial parameter controlling the accreting fate of the system: the distribution of impact velocities $dv$ amongst planetesimals orbiting HD196885A. Due to stringent computing time constraints, we adopt a deterministic model based on several simplifications (static gas disc, no disc gravity, etc.),  all of which concurring to give a conservative lower limit for the $dv$ distribution.
 
We find that, for almost the whole explored parameter space (planetesimal sizes, gas disc profile, inclination between the binary and the circumprimary plane, etc.), impact velocities reach values that lead to eroding impacts between planetesimals. Most of the circumprimary disc is thus strongly hostile to accretion, especially the region at 2.6\,AU from the primary corresponding to the current location of the planet. 

We considered the possibility that the planet was formed in an accretion-friendly region much closer to the primary and later moved outward, either during the embryo accretion phase or by mutual perturbations with another (undetected) giant planet. However, we rule out this hypothesis because the inner accretion-safe region is much too narrow and close to the primary, $r\leq 0.4\,$AU, to allow the formation of two giant planets.

In the highly perturbed $r\sim 2.6\,$AU region, the only way for the HD196885Ab planet to form through the mutual accretion of planetesimals is either 1) if these planetesimals were initially larger than $\sim 250\,$km, or 2) if the binary had an initial separation $a_b \geq 45\,$AU and shrinked by at least a factor 2 during its early history. Although large initial planetesimals as well as early orbital compacting of binaries both make sense in view of recent planetesimal-formation and stellar cluster theories, the values we find for these 2 parameters do appear rather extreme.
An alternative solution is to suppose that planetesimal grow through the so-called "snowball" growth mode \citep{xie10b} by progressively sweeping up small dust particles. However, the efficiency of this alternative growth channel has not been quantitatively estimated yet, especially in the specific context of close binaries. 

These results strengthen HD196885Ab's status as the most "extreme" planet-in-a-binary known to date. What sets it appart from other cases, like the $\gamma$ Cephei planet or a putative habitable planet in the $\alpha$ Cen system, for which the canonical core-accretion scenario also encounters serious problems, is that it is the only one for which these problems could be serious enough to become insurmontable. 
Should this be the case, then alternative planet-formation scenarios, such as direct collapse due to disc instability, might be considered for this planet, and possibly all other giant planets in close binaries \citep{duch10}. Unfortunately, the disc instability scenario does also encounter severe difficulties in the context of close binaries, and it is too early to know it can be a viable alternative formation channel.

\begin{acknowledgements}

We thank Gael Chauvin for enlightening discussions and comments. We also thank both anonymous reviewers whose comments considerably helped to improve the paper.

\end{acknowledgements}

\end{document}